\documentclass[12pt]{article}
\pdfoutput=1
\usepackage{graphicx,amssymb,amsmath}
\usepackage[sort&compress,numbers]{natbib}
\graphicspath{{figs/}}
\usepackage[colorlinks=true,urlcolor=blue,
anchorcolor=blue,citecolor=blue,filecolor=blue,
linkcolor=blue,menucolor=blue,
linktocpage=true,pdfa=true]{hyperref}

\usepackage[left=2cm, right=2cm, top=1.9cm, bottom=0.75cm,
footskip=48pt, headsep=20pt, includehead, includefoot]{geometry}

\newcommand{\fig}[1]{\includegraphics[width=.48\textwidth]{#1}}
\newcommand{\figw}[2]{\includegraphics[width=#1\textwidth]{#2}}
\newcommand{\figh}[2]{\includegraphics[height=#1\textheight]{#2}}
\newcommand{\uhecrs}{\mbox{UHECRs}}

\begin{document}
\title{\bfseries
	Application of neural networks to classification of data
	of the TUS orbital telescope}
\author{Mikhail Zotov\\
	\itshape D.V. Skobeltsyn Institute of Nuclear Physics\\
	\itshape M.V. Lomonosov Moscow State University\\
	\itshape Moscow, Russia}
\date{}
\maketitle
\begin{abstract}

	We employ neural networks for classification of data of the TUS
	fluorescence telescope, the world's first orbital detector of
	ultra-high energy cosmic rays. We focus on two particular types of
	signals in the TUS data: track-like flashes produced by cosmic ray
	hits of the photodetector and flashes that originated from distant
	lightnings. We demonstrate that even simple neural networks combined
	with certain conventional methods of data analysis can be highly
	effective in tasks of classification of data of fluorescence
	telescopes.

\end{abstract}

\tableofcontents

%______________________________________________________________________________
\section{Introduction}

It is well known that charged particles in an extensive air shower
generated by a cosmic ray hitting the Earth's atmosphere interact with
atmospheric nitrogen, causing it to emit ultraviolet (UV) light via a
process called fluorescence. Registration of this UV emission in the
nocturnal atmosphere with dedicated telescopes is an established
technique of studies in the field of ultra-high energy cosmic rays
(\uhecrs) that demonstrated its efficacy with the Fly's Eye detector
array in 1980--1993~\cite{FlysEye}.  Nowadays fluorescence telescopes
are an inherent part of both main modern experiments dedicated to such
studies---the Pierre Auger Observatory~\cite{Auger} and the Telescope
Array~\cite{TA} complementing their surface detectors.

One of the main problems in UHECR studies is their extremely low flux at
energies $\gtrsim5\times10^{19}$~eV, above the so-called
Greisen--Zatsepin--Kuzmin cut-off~\cite{Greisen-1966,ZK-1966}. This necessitates
running ground experiments occupying huge areas. For example, the Pierre
Auger Observatory covers an area of $\sim3000~\text{km}^2$. In early
1980's, Benson and Linsley suggested to increase dramatically the
exposure of UHECR experiments by a totally different approach.  They
suggested to put a wide-field-of view fluorescence telescope into a
low-Earth orbit and employ it for registering fluorescence and Cherenkov
emission of extensive air showers born by \uhecrs{} in the nocturnal
atmosphere~\cite{BL80,BL81}.

Several projects aimed to implement the~idea of Benson and~Linsley have
been suggested since then, among them Orbiting Wide-field
Light-collectors (OWL)~\cite{OWL},  Tracking Ultraviolet Setup (TUS) and
KLYPVE (Kosmicheskie Luchi Predel'no Vysokikh Energii, which stands for
Extreme Energy Cosmic Rays in Russian)~\cite{2001AIPC..566...57K,
2001ICRC....2..831A}, the former Extreme Universe Space Observatory on
board the Japanese Experiment Module, now Joint Experiment Missions for
Extreme Universe Space Observatory (JEM-EUSO)~\cite{JEM-EUSO}
and~others.  TUS was the first one launched into space in
2016~\cite{JCAP2017}.  It was a comparatively simple device designed as
a prototype of the KLYPVE project. Its main aim was to test the
technique suggested by Benson and Linsley.  The data set obtained with
TUS has demonstrated that an unexpected variety of different processes
are taking place in the nocturnal atmosphere of the Earth in the
ultraviolet (UV) range.  Considerable efforts have been put in the
analysis of the TUS data, see, e.g.,~\cite{UHECR2016,
2019EPJWC.21006006B, 2019RemS...11.2449K, JCAP2020,
2020CosRe..58..317K}.  Until recently, all of them were based on
conventional approaches.  Results of the first attempt of classifying
the TUS data with neural networks were briefly presented
in~\cite{BMY2020}. The work was motivated by a number of reasons.  On
the one hand, TUS registered more than 78 thousand events in the main
mode of operation aimed at observing the fastest phenomena in the
atmosphere, and more than almost 12 thousand events in other modes, with
every event represented by 65,536 numbers. A visual analysis of this
data set is a time-taking and error-prone process, which must be
supported by established methods of data analysis. On the other hand,
methods that involve machine learning and neural networks are ubiquitous
nowadays in many branches of science and everyday life but their
application in the field of cosmic ray physics is still rather limited,
see a few examples in~\cite{2018APh....97...46E, 2019APh...111...12G,
2019ICRC...36..456V, 2020JPhCS1525a2001K, 2020arXiv200507117I,
2021APh...12902579S, 2021arXiv210102946T} and~\cite{2020JPhCS1525a2112E}
for a brief review.

The Mini-EUSO fluorescence telescope is currently working onboard the
International Space Station~\cite{mini-EUSO}, a stratospheric experiment
EUSO-SPB2 equipped with a fluorescence and a Cherenkov telescopes is to
be launched in 2023~\cite{EUSO-SPB2} and an advanced POEMMA
(Probe Of Extreme Multi-Messenger Astrophysics) mission
consisting of two orbital telescopes aimed at registering UHE cosmic
rays and neutrinos is being developed~\cite{POEMMA}. The spatial
resolution and the amount of data of these experiments is expected to be
orders of magnitude bigger than that of TUS. This suggests developing
methods suitable for an analysis of their data, including methods based
on neural networks. The work~\cite{BMY2020} was the first step in this
direction applied to the TUS data.  In the present paper, we further
develop the approach suggested in~\cite{BMY2020} and present more
results of applying neural networks to classification of data of the TUS
mission.

%______________________________________________________________________________
\section{The TUS detector}\label{sec:TUS}

A comprehensive description of the TUS telescope can be found
in~\cite{SSR2017,JCAP2017}.
Here we will briefly outline its main features important for the article.

From the very beginning TUS, was designed as a simple device, a
pathfinder for a much more sophisticated KLYPVE experiment onboard the
International Space
Station~\cite{2001AIPC..566...57K,2001ICRC....2..831A}.  The main
components of TUS were a Fresnel mirror and a square-shaped $16\times16$
photodetector aligned to the focal surface of the mirror.  The mirror
had an area of nearly 2~m$^2$ and a 1.5~m focal distance.  The field of
view (FOV) of the telescope was $9^\circ\times9^\circ$, which covered an
area of approximately $80~\text{km}\times80~\text{km}$ at sea level. Two
hundred fifty-six Hamamatsu R1463 photomultiplier tubes (PMTs) formed
the focal surface of the photodetector.
A glass UV filter was placed in front of every PMT to limit the measured
wavelength to the 300--400~nm range.  Light guides with square entrance
apertures and circular outputs were used to uniformly fill the FOV. All
pixels had 1~cm black blends to protect them from side illumination but
the telescope as a whole did not have any side shields.

All pixels were grouped in 16 modules, each of which had its own
high-voltage power supply and a data processing system for the
first-level trigger. The high-voltage system (HVS) was aimed to adjust the
sensitivity of PMTs to the intensity of the incoming light and switch
them off completely on day sides of the orbit.

The TUS electronics could operate in four modes with different time
sampling windows. The main mode was intended for registering the fastest
processes in the atmosphere and had the time step of~0.8~$\mu$s. Every
record consisted of ADC codes written for all photodetector channels in
256 time steps with a total duration of 204.8~$\mu$s. We only discuss
events registered in this mode of operation.

\begin{figure}[!ht]
	\centering
	\fig{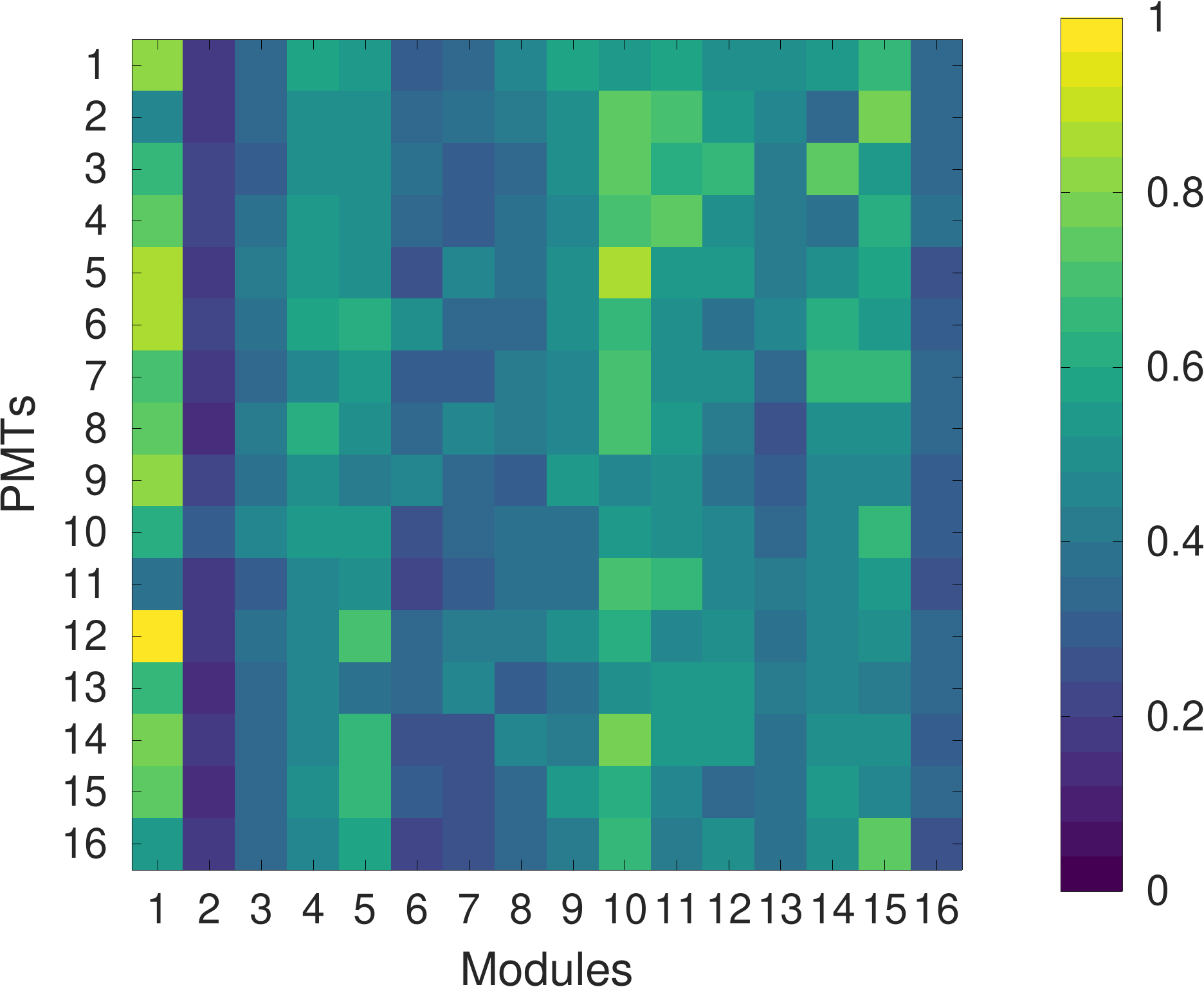}\quad\fig{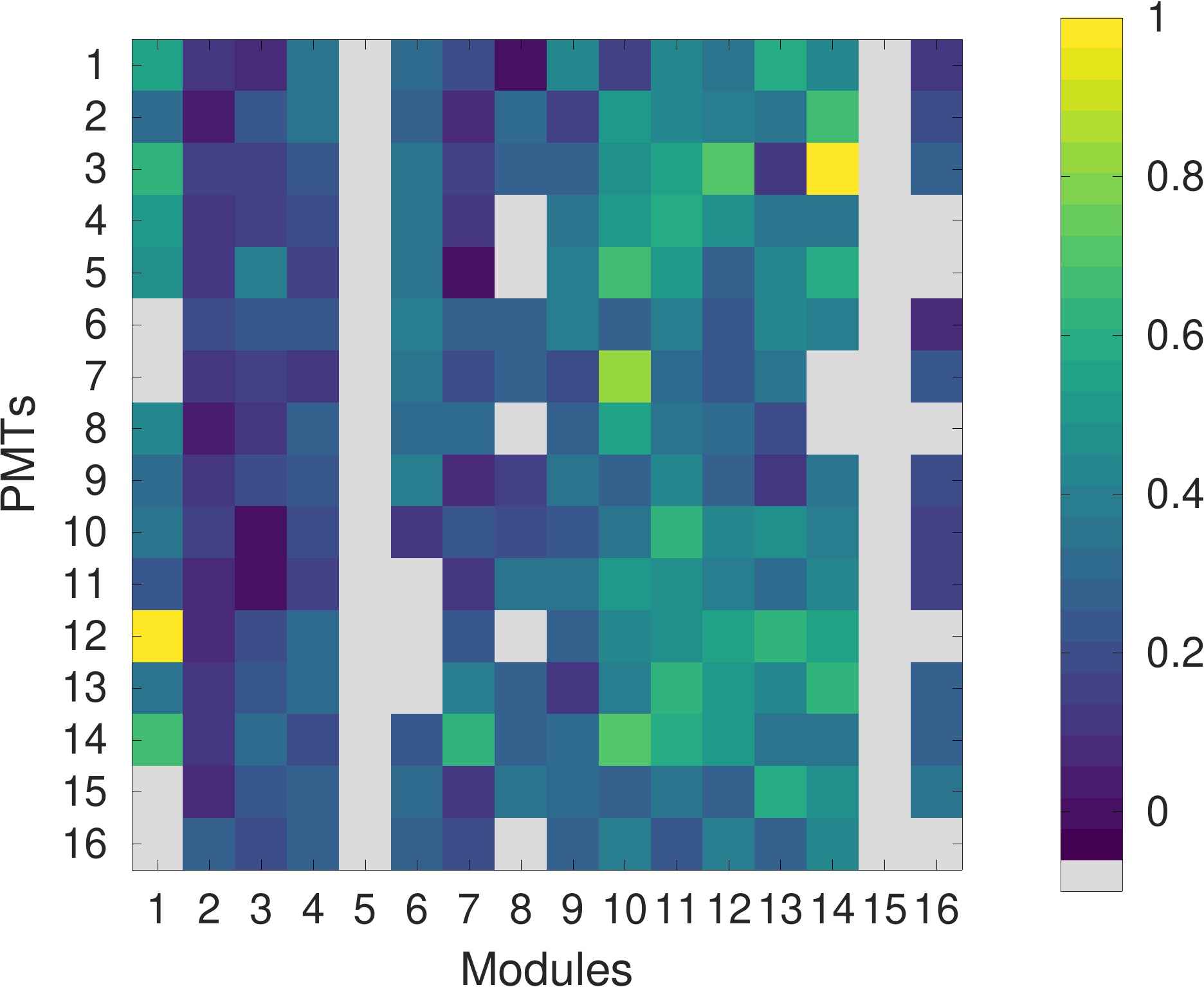}

	\caption{Relative sensitivities of channels of the TUS photodetector.
	Left: pre-flight measurements (adapted from~\cite{SSR2017}).
	Right: estimations obtained from the data. Dead channels are shown
	in grey.}
	
	\label{pmt_gains}
\end{figure}

TUS was launched into orbit on April~28, 2016, as a part of the
scientific payload of the Lomonosov satellite. The satellite had a
sun-synchronous orbit with an inclination of $97.3^\circ$, a height
about 470--515~km and a period of $\approx94$ minutes.  The detector was
first put into operation on May~19, 2016.  Unfortunately, a shortcut
took place in the HVS of the photodetector at the very first day of
operation~\cite{JCAP2020}.  Due to this, the device operated at the
highest voltage during the first few orbits, including day segments,
until information about this arrived to the operating center and the
detector was switched off.  As a result of the accident, the HVSs of two
modules were burnt and these modules did not operate anymore.  In other
modules, PMTs with a high current were damaged.  Totally, 51
photodetector channels were burnt. They are shown in grey in the right
panel of Figure~\ref{pmt_gains}. Besides this, sensitivities of other
channels changed.

A number of attempts to perform an in-flight calibration of the
photodetector have been made~\cite{JCAP2020, Pavel-PMTgains}.  These
methods, as well as a method tested during the present work (see
subsection~\ref{subsec:slow_flashes}), resulted in estimates of relative
sensitivities that qualitatively agreed with each other for the majority
of channels. Figure~\ref{pmt_gains} shows relative sensitivities of
photodetector channels calculated from the pre-flight calibration and
estimated after the mission. However, none of the approaches has
provided reliable estimations of absolute PMT gains. More than this, all
estimates of relative sensitivities of photodetector channels were
obtained for the highest codes of the HVS, which correspond to perfect
observational conditions, but only 38.4\% of data satisfy this
criterion.  Together with a few changes of the trigger settings made
during the mission, this poses certain problems for the data analysis
within a unified approach.

Regular operation of TUS began on August~18, 2016, and continued till
November~30, 2017.

%______________________________________________________________________________
\section{Phenomenological classification of the TUS data}

A preliminary phenomenological classification of TUS events was ready
by October 2016~\cite{UHECR2016} and was refined later
on~\cite{JCAP2017}.
All events were divided into several groups:
\begin{itemize}

	\item events with stationary noise-like waveforms; these included
		events with a strongly non-uniform illumination of the focal
		surface with bright regions correlated with geographical positions
		of cities and objects like airports, power plants, offshore
		platforms etc.

	\item instant track-like flashes (TLFs) caused by charged particles
		hitting the UV filters of the photodetector;

	\item flashes produced by light coming outside of the FOV of the
		detector and scattered on its mirror; they were called ``slow
		flashes'' because of the long signal rise time in comparison
		with TLFs; and

	\item events with complex spatio-temporal dynamics; these included
		so called ELVEs, which are short-lived
		optical events that manifest at the lower edge of the
		ionosphere (altitudes of $\sim90$~km) as bright rings expanding
		at the speed of light up to a maximum radius of
		$\sim300$~km~\cite{1996GeoRL..23.2157F},
		events with waveforms that could be expected from fluorescence
		originating from extensive air showers produced by
		extreme energy cosmic rays, as well as violent flashes
		of a yet unknown origin.

\end{itemize}

Miscellaneous examples of events of all types can be found
in~\cite{JCAP2017, UHECR2016, 2019EPJWC.21006006B, 2019RemS...11.2449K,
JCAP2020,2020CosRe..58..317K}.
Note that the four classes of events were not disjoint: in some cases
an event of one type was registered simultaneously with an event of
another type within one data record.
In what follows, we shall focus on instant track-like flashes and slow
flashes. Let us recall their main features before discussing an
application of neural networks for their search.

\subsection{Instant track-like flashes}

\begin{figure}[!ht]
	\centering
	\fig{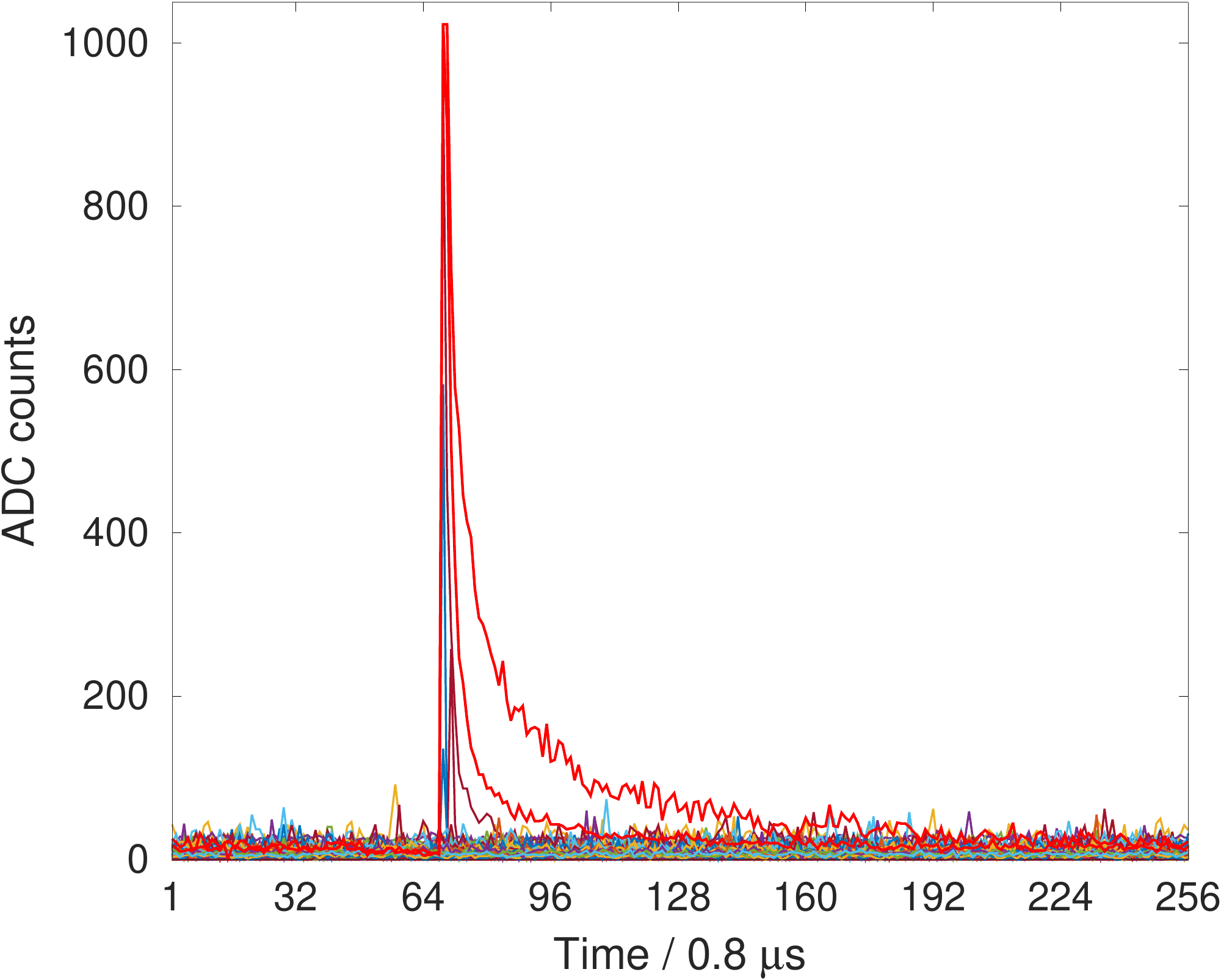}\quad\fig{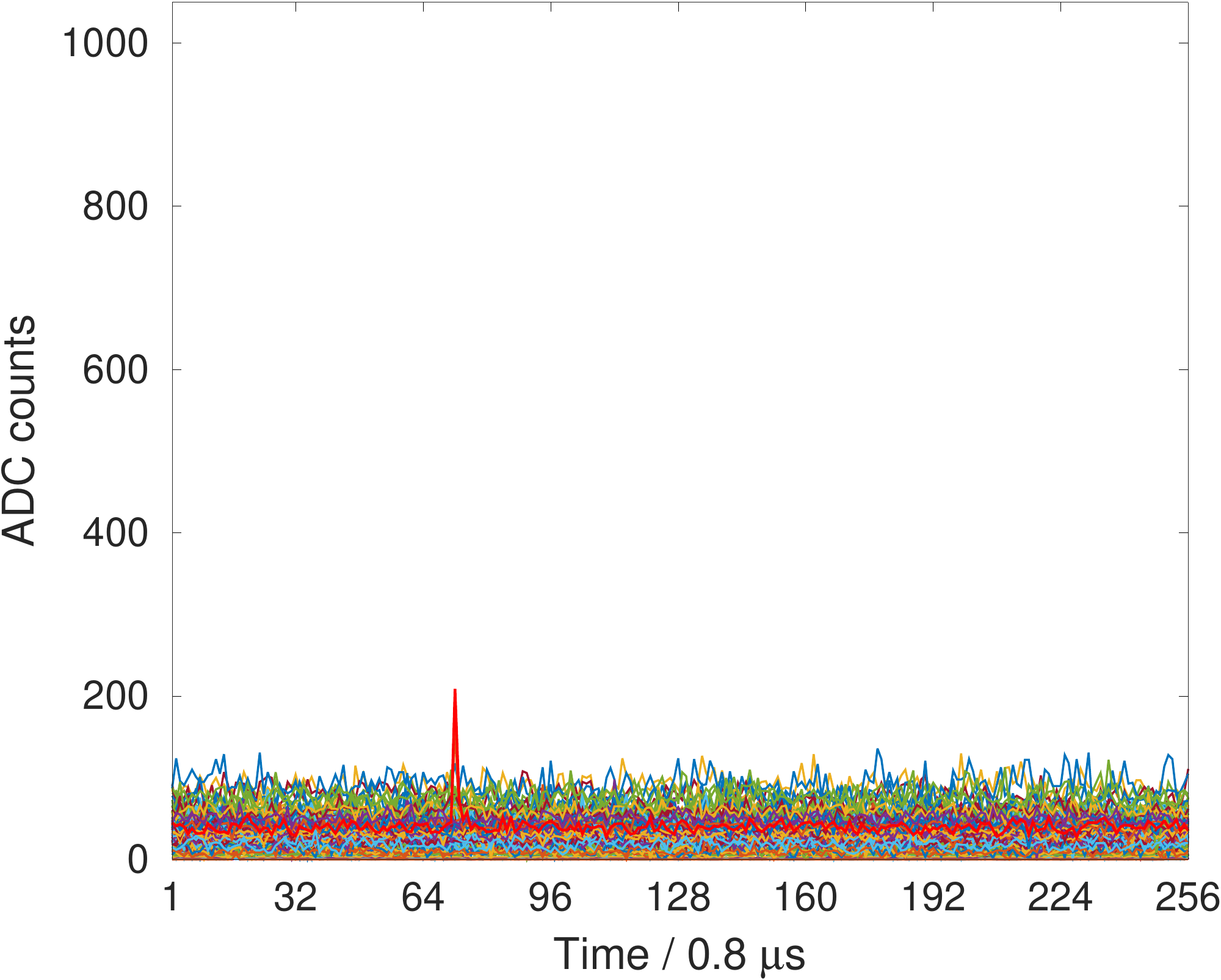}
	\caption{Examples of waveforms of instant track-like flashes.
	Left: typical waveforms with ADC codes reaching their maximum values
	in one or more channels (shown in red). Right: a weak TLF on a strong
	background registered during a full-Moon night.
	One time tick equals 0.8~$\mu$s, the total duration of a record
	is 204.8~$\mu$s.}
	\label{tracks-waveforms}
\end{figure}

TLFs were observed in the TUS data from the very beginning of the
mission on May~19, 2016.  They demonstrated an instant growth of the
signal (in one or two time steps), mostly to the highest possible ADC
codes, in a single pixel or a group of pixels, sometimes producing a
kind of a linear track on the focal surface.  This is why they were
called ``instant track-like flashes.'' They comprised about 10\% in
average of all events and up to 25\% during moonless nights before they
were suppressed in late April, 2017, with an update of the onboard
software.  A typical shape of a TLF is shown on the left panel of
Figure~\ref{tracks-waveforms}. However, their signals were dim in some
cases as demonstrated on the right panel of
Figure~\ref{tracks-waveforms}. 

\begin{figure}[!ht]
	\centering
	\fig{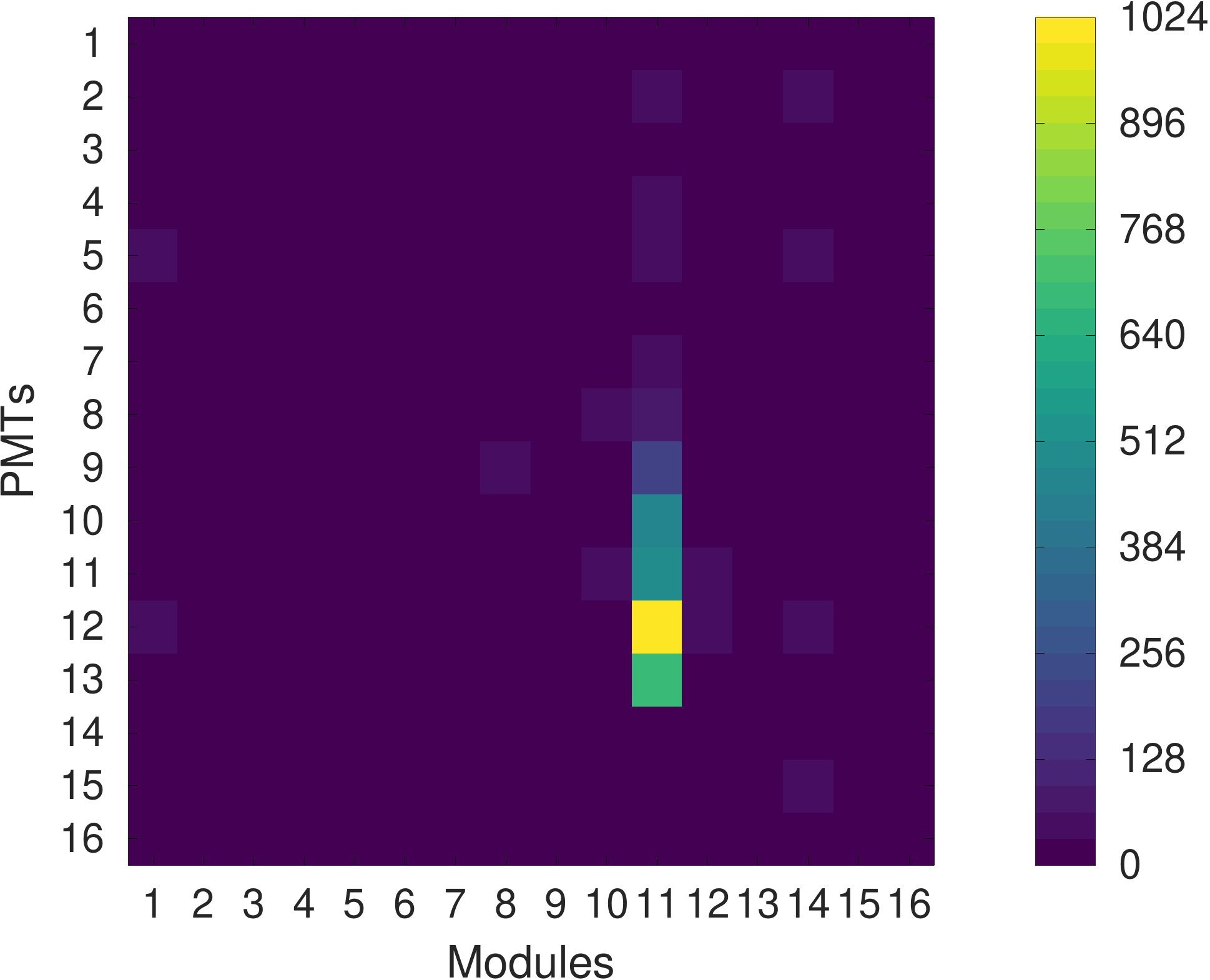}\quad\fig{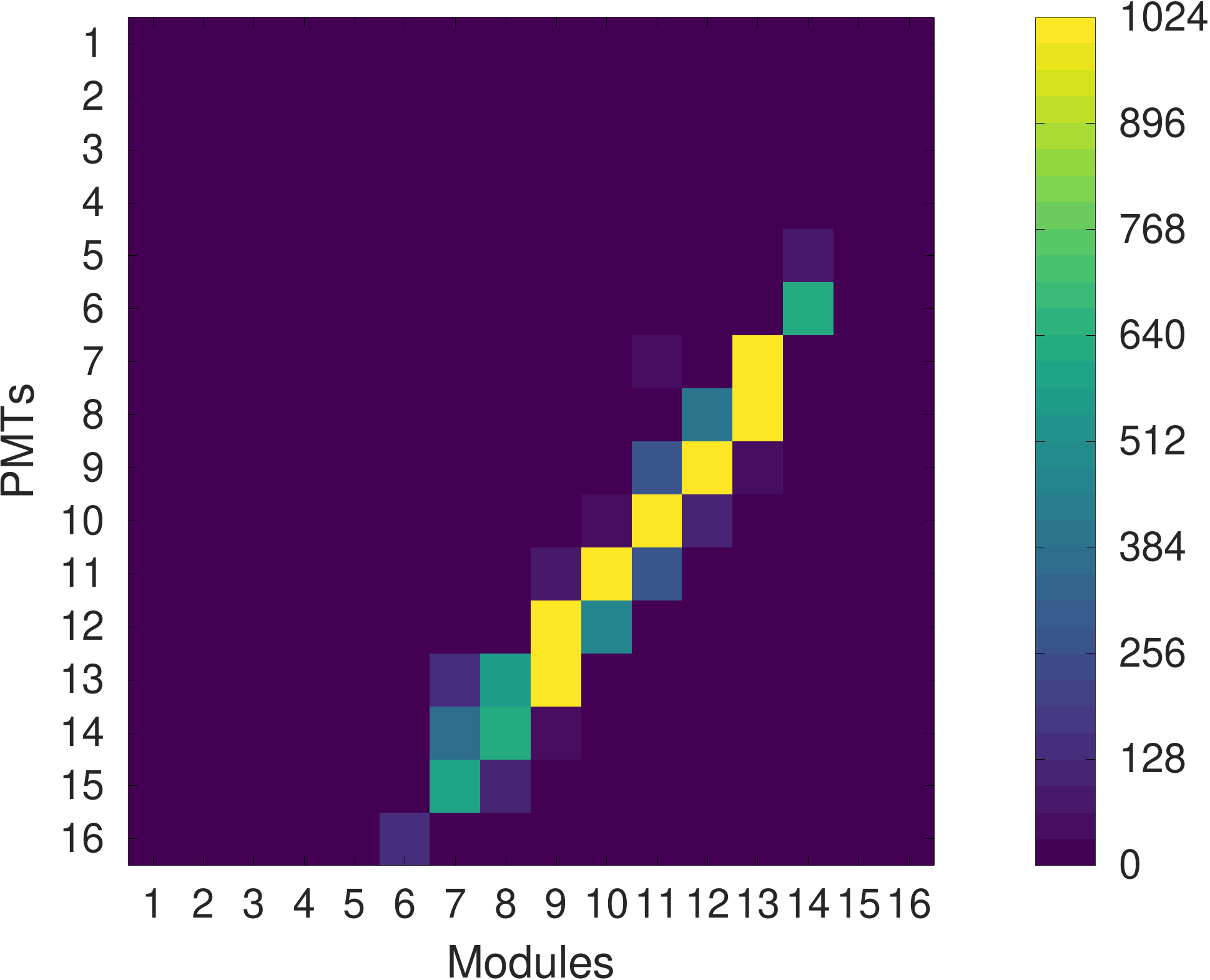}
	\caption{Examples of ``snapshots'' of the focal surface at the moment
	of the (first) maximum of TLFs with clear linear tracks.
	Here and in what follows, colours in images like these denote the
	value of ADC counts as shown in the colour bars.}
	\label{tracks-nice}
\end{figure}

\begin{figure}[!ht]
	\centering
	\fig{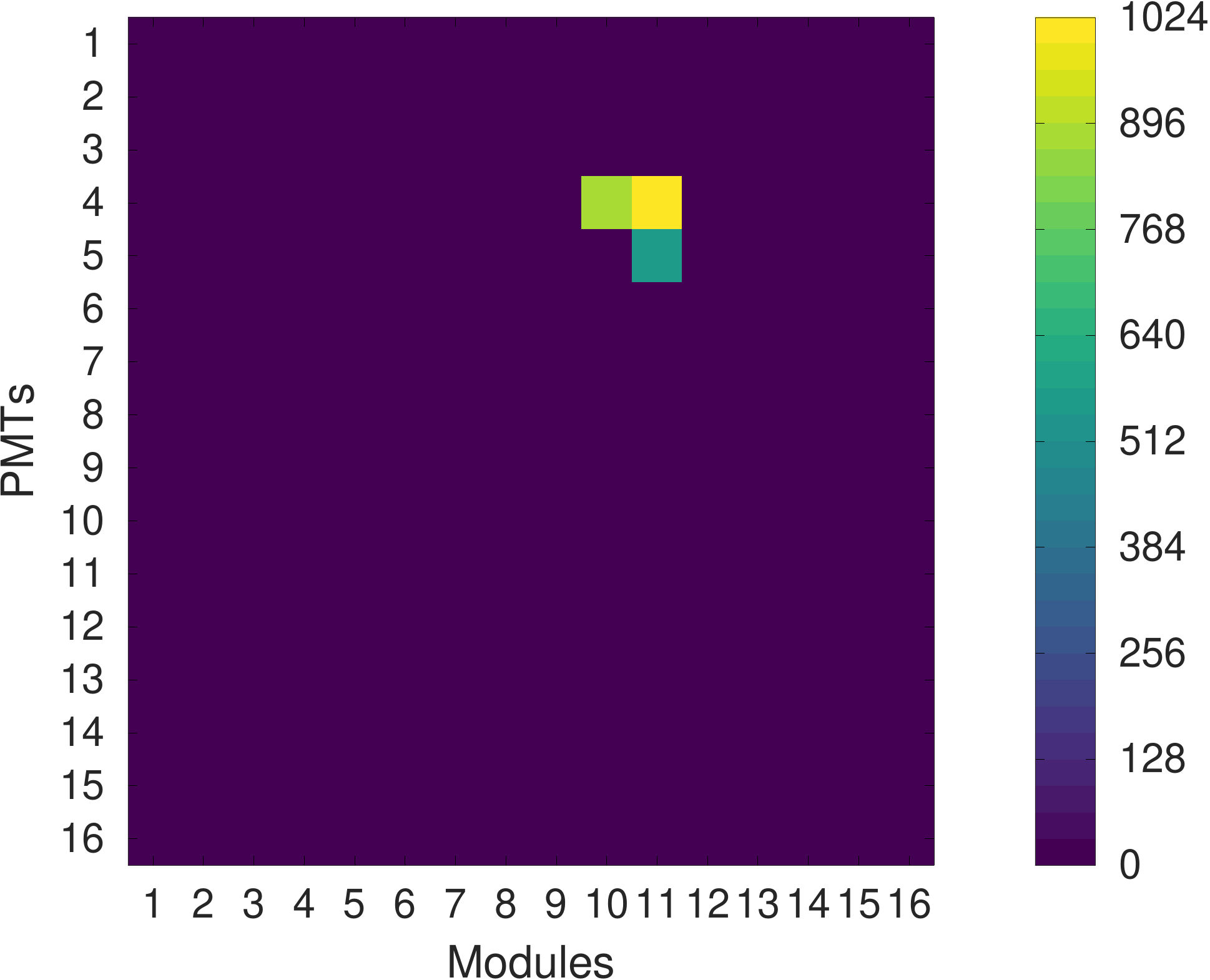}\quad\fig{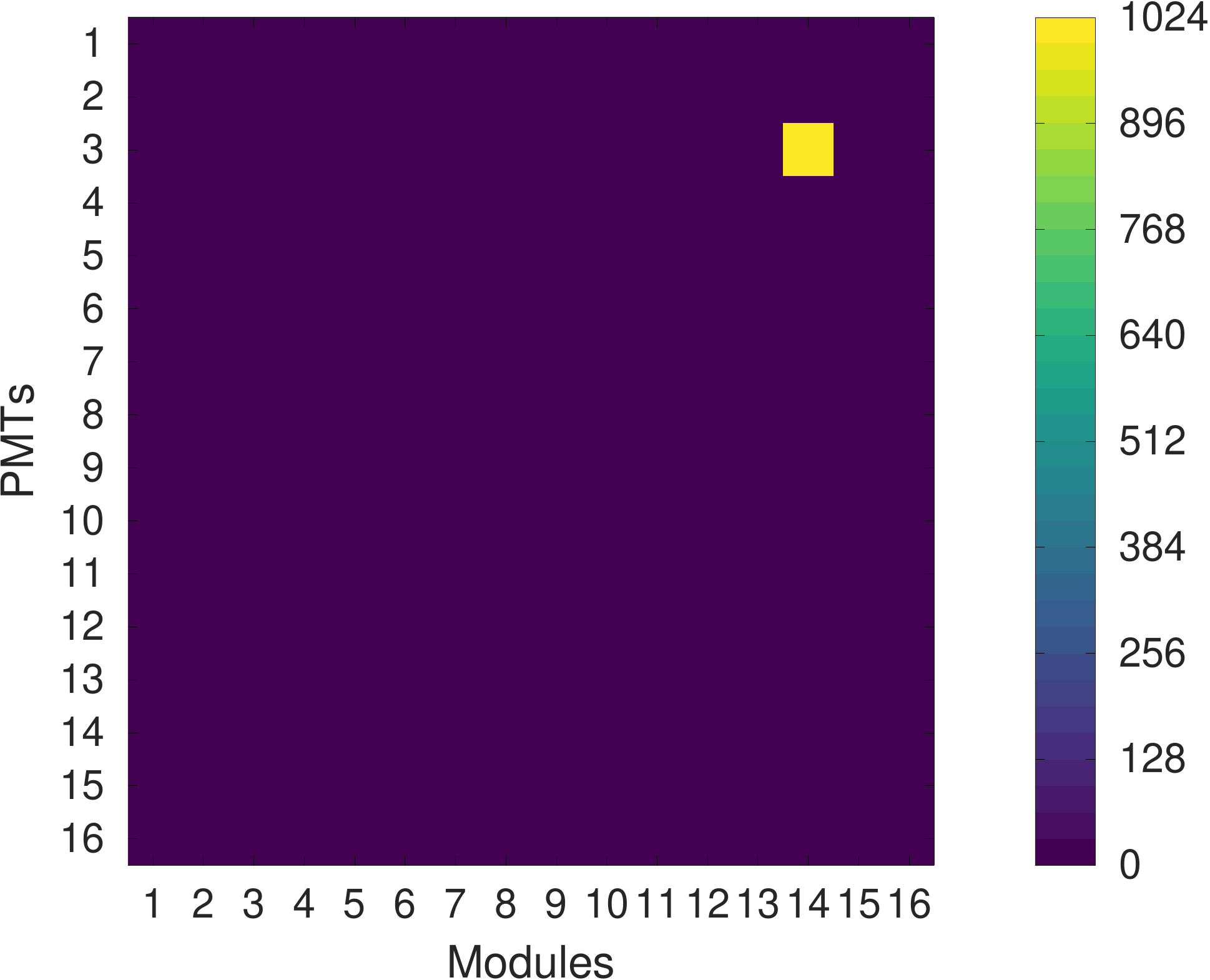}
	\caption{Examples of traces of TLFs that made active only a small
	group or even a single channel.}
	\label{tracks-spot}
\end{figure}

\begin{figure}[!ht]
	\centering
	\fig{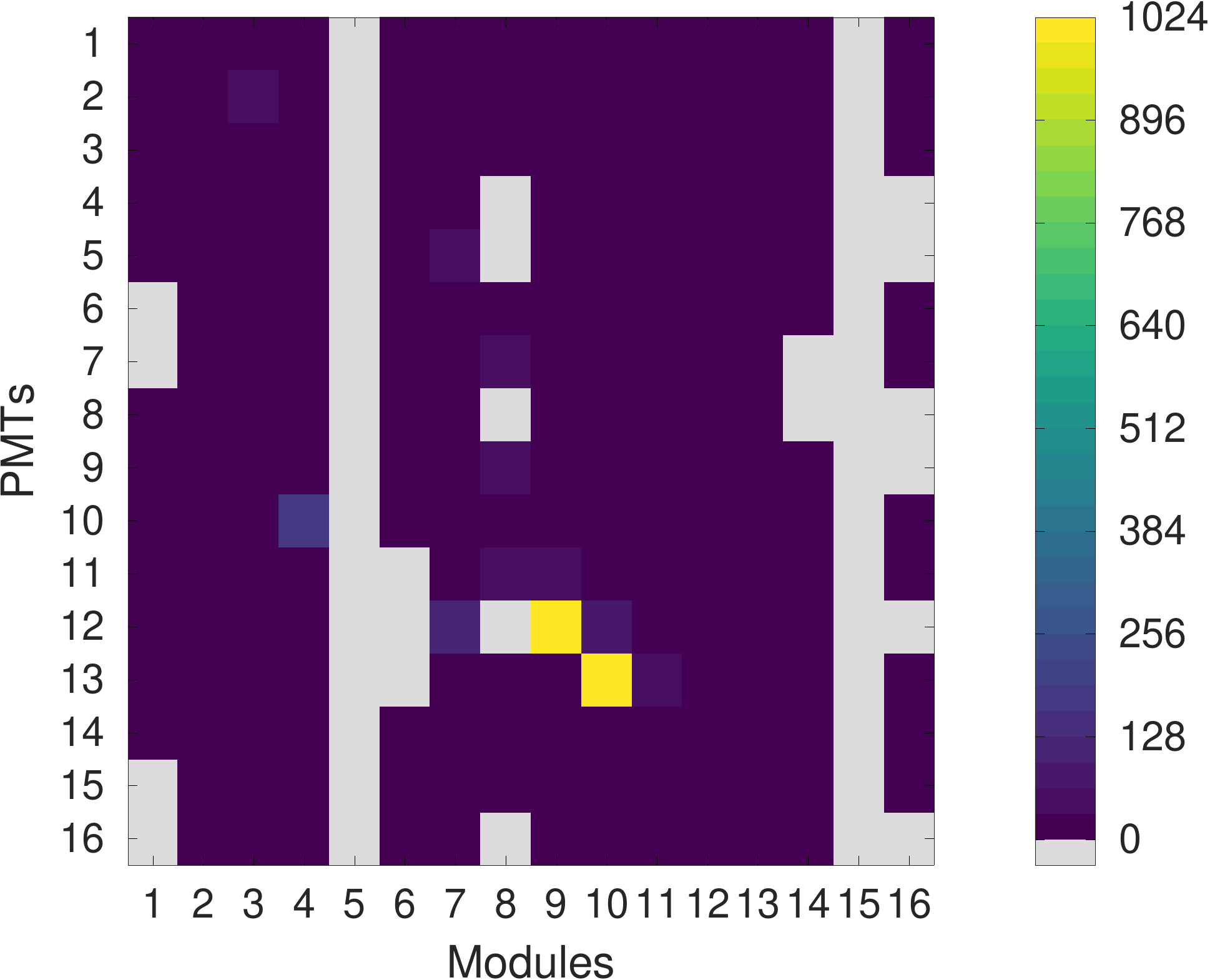}\quad\fig{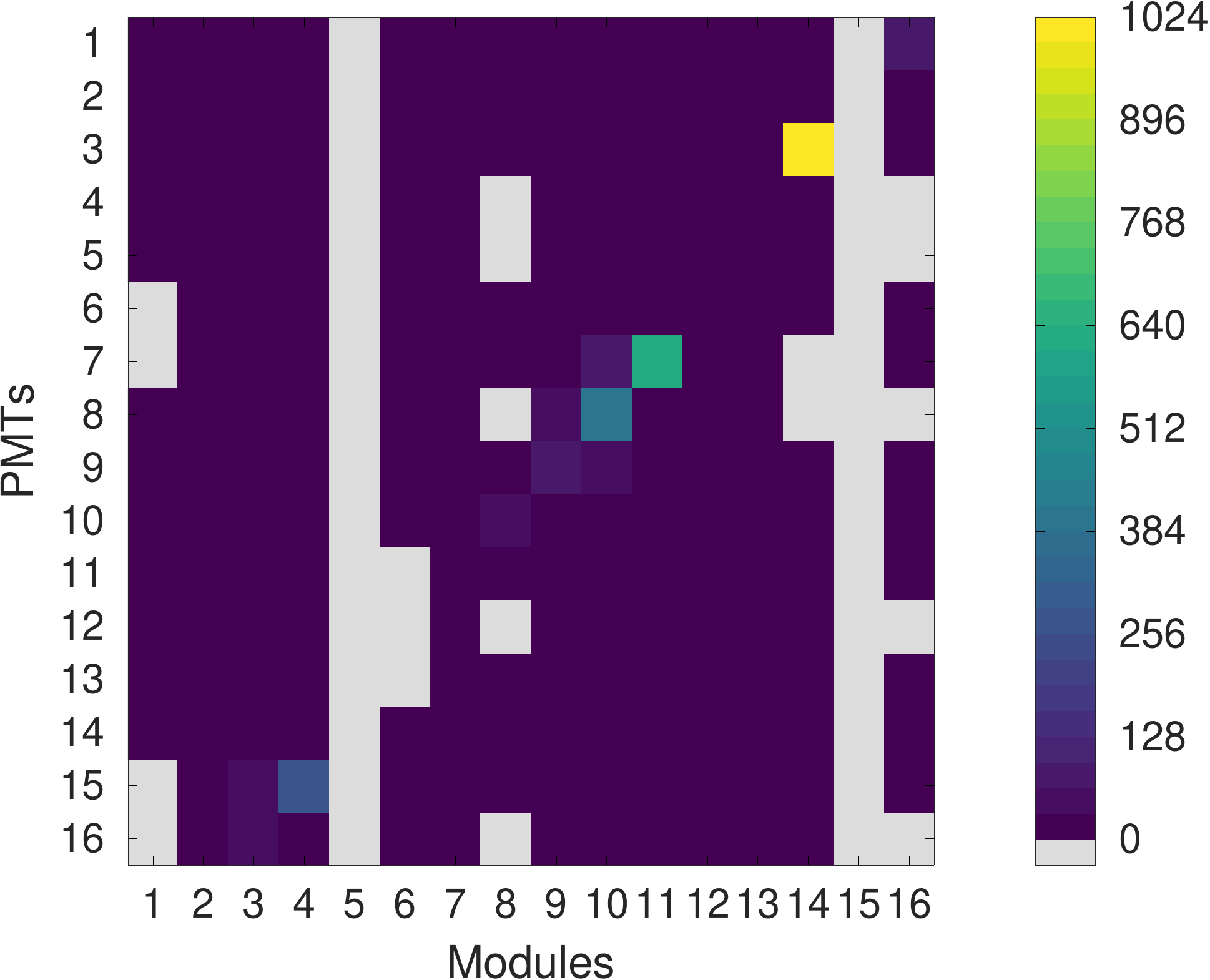}
	\caption{Examples of traces of TLFs ``broken'' by intersecting
	dead channels, see the text for details.}
	\label{tracks-broken}
\end{figure}

\begin{figure}[!ht]
	\centering
	\fig{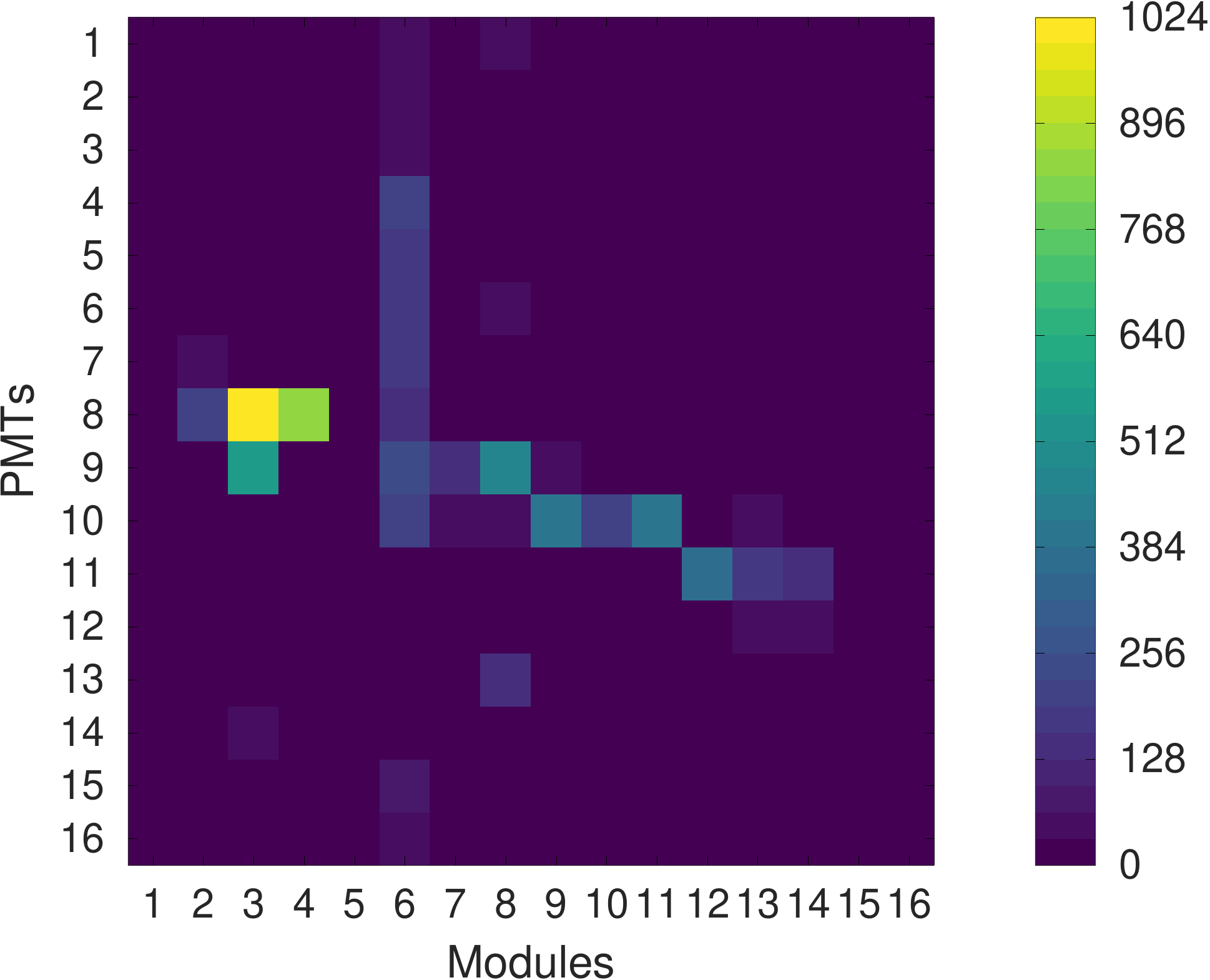}\quad\fig{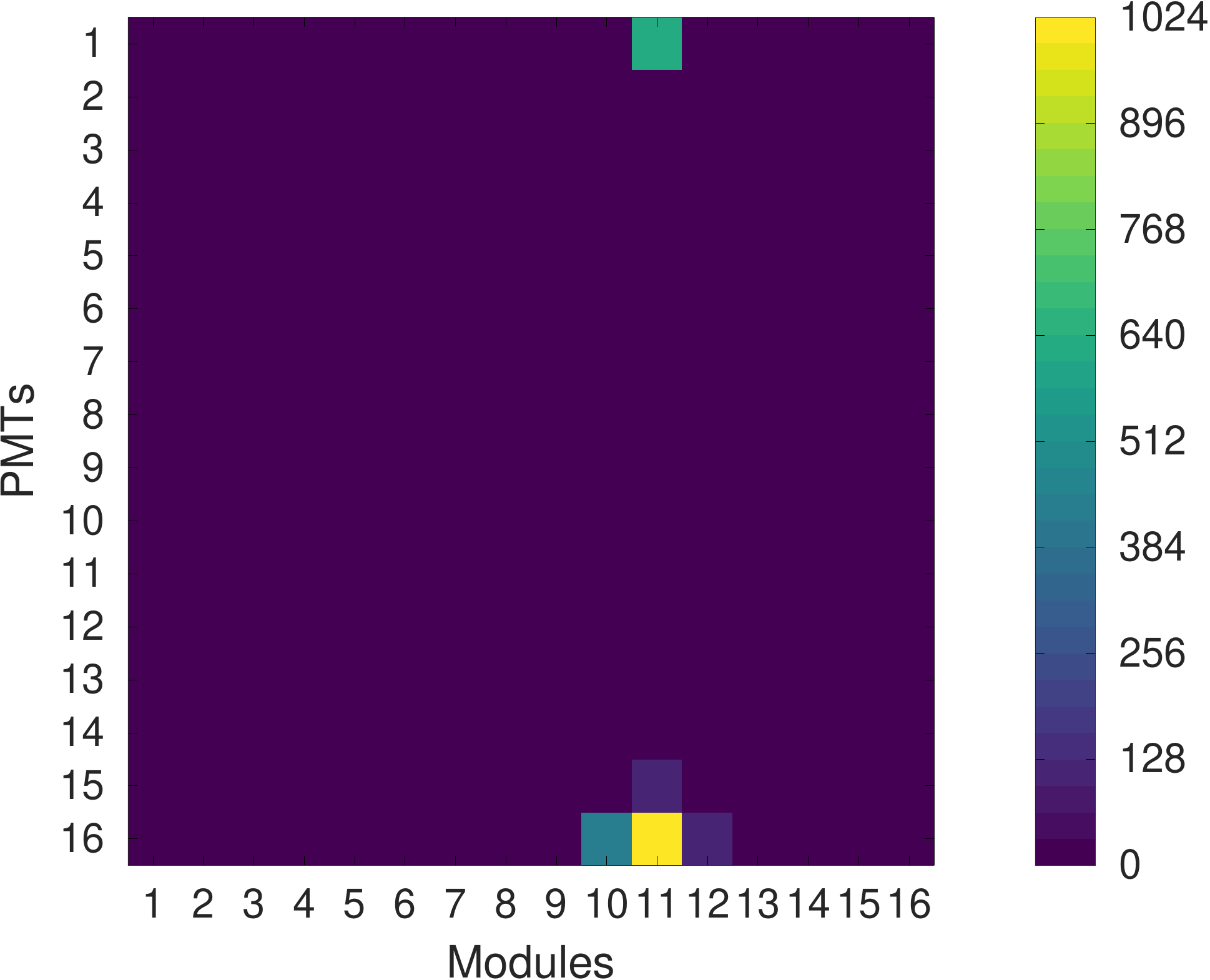}
	\caption{Left: an examples of an X-shaped trace consisting of a TLFs
		trace and a simultaneous crosstalk in module~6. Right: a short
		track at the bottom edge of the focal surface and a simultaneous
		signal on the other end of module~11 due to a feature of the
		electronics operation.}
	\label{tracks-electro}
\end{figure}

While waveforms of TLFs could be easily selected by eye in the majority
of cases, their traces on the focal surface were much more diverse.
Figure~\ref{tracks-nice} provides a couple of examples of events in
which clear linear tracks could be seen in the focal surface.
In other cases, the traces of TLFs were not that evident. A few typical
examples are provided in
Figs.~\ref{tracks-spot}, \ref{tracks-broken} and~\ref{tracks-electro}.

In many cases, TLFs produced a signal in a small group or even in a
single channel of the focal surface creating a spot-like trace similar
those shown in Figure~\ref{tracks-spot}.  This could take place both
for weak flashes with a low amplitude and for flashes that saturated the
channels as shown in the figure.

Figure~\ref{tracks-broken} presents two examples of linear traces broken
by passing through inactive or low-sensitivity channels.  In the left
panel, a track spanning through seven modules is clearly broken when
intersecting a number of dead channels. The situation is more complex
for the event shown in the right panel.  The track spans from PMT~1 in
module~16 to PMT~16 in module~3.  It crosses two dead modules, a group
of dead PMTs in module~6 and a number of low-sensitivity channels in
modules 7, 8 and~9, cf.\ the right panel of Figure~\ref{pmt_gains}.
Besides this, the instant growth of the signal in modules~12 and~13 is
recorded one time step later due to the way ADC codes were read out. As
a result, we obtain a ``dashed'' track.  That was a typical situation
since 51 channels were dead (i.e., almost 20\% of the focal surface),
and PMT gains of a whole number of channels decreased drastically after
the misbehaviour of the HV system discussed in Section~\ref{sec:TUS}.

Figure~\ref{tracks-electro} provides two examples of situations when a
signal of a TLF was distorted due to some peculiarities in the work of
onboard electronics.  An X-shaped trace shown in the left panel
originated because of a crosstalk in module~6.  A situation like this
used to occur mostly with module~6 but could also be found in a less
pronounced form with module~8.  The right panel of
Figure~\ref{tracks-electro} presents an example of a TLF that took place
at the edge of the focal surface at one end of a module and produced a
simultaneous signal in a channel at the opposite side of the active
module.

Understanding the nature of TLFs was straightforward since long tracks
in the focal surface appearing in less than a microsecond could not be
produced by flashes in the atmosphere. A simulation performed with
GEANT4~\cite{Geant4} confirmed that such signals could be produced by protons
with energies in the range from a few hundred MeV to several GeV hitting
the photodetector (mostly the UV filters)~\cite{IzvRan2017}.
An analysis of the latitudinal distribution of TLFs also supported this
hypothesis~\cite{JCAP2017}.

TLFs presented a class of ``parasitic'' events that occupied a part of
active time of the detector. An anti-trigger was implemented in the
onboard software on April~28, 2017, which effectively reduced their
number in the TUS data by an order of magnitude.

%______________________________________________________________________________
\subsection{Slow flashes}
\label{subsec:slow_flashes}

As it was already mentioned above, slow flashes were named as such
due to the long signal rise time in comparison with TLFs. Similar to
TLFs, they are ubiquitous in the TUS data set.

\begin{figure}[!ht]
	\centering
	\fig{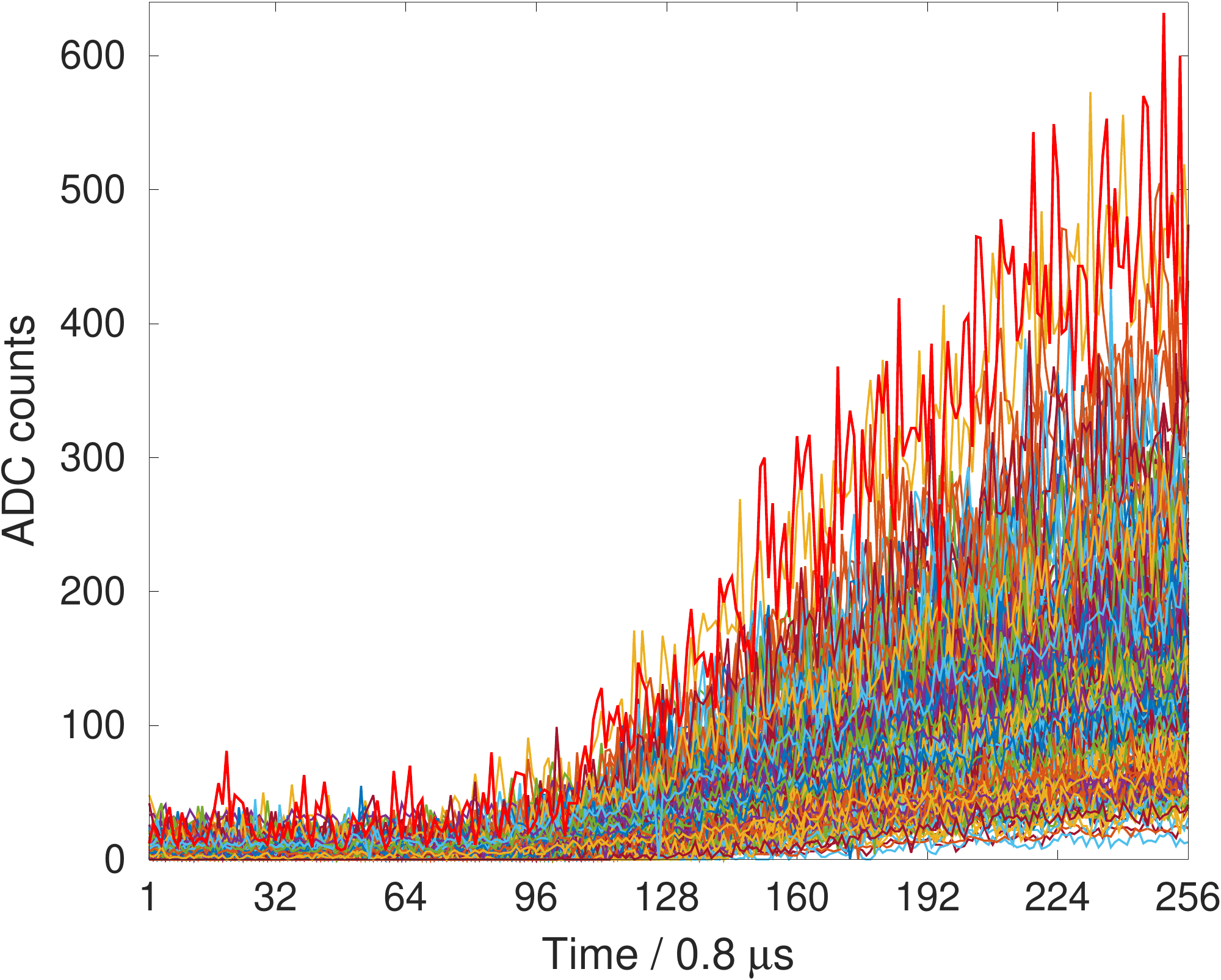}\quad\fig{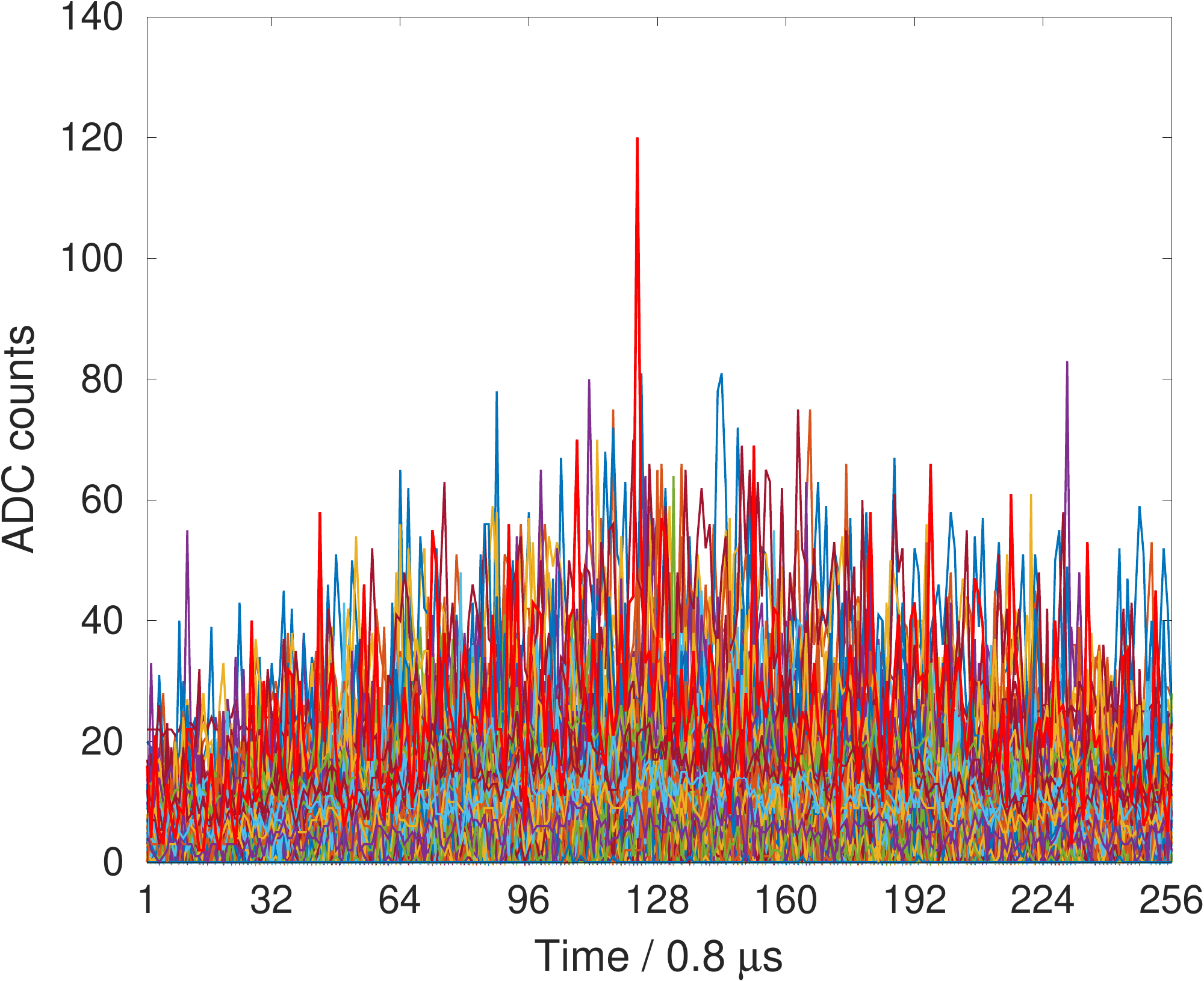}
	\caption{Two types of waveforms typical for slow flashes.}
	\label{sf-typical}
\end{figure}

All slow flashes can be roughly divided into two groups: longer flashes
that do not reach their maximum during a TUS record (see the left panel
of Figure~\ref{sf-typical}) and shorter flashes in which the signal
passes its maximum and then attenuates, sometimes even reaching the
level preceding the flash (see the right panel of
Figure~\ref{sf-typical}).  Slow flashes of both groups differ in their
amplitudes and the exact shape of waveforms but illumination of the
focal surface is practically uniform in most cases up to sensitivity of
channels. During this work, we have selected 113 slow flashes in which a
rise of the signal was observed in at least 196 channels at
$\ge10\sigma$ and calculated relative sensitivities of active channels
assuming uniform illumination of the focal surface. The results
qualitatively agreed with estimates obtained with a totally different
technique presented in~\cite{Pavel-PMTgains}.

A comparison of geographic locations of slow flashes gave an immediate
clue to their origin: they demonstrated a clear correlation with regions
of high thunderstorm activity~\cite{JCAP2017}. A set of then-known
strong slow flashes registered from August 16, 2016, till September, 19,
2016, was compared with the data on lightning strikes kindly provided
by the World-Wide Lightning Location Network (WWLLN). It was found that
there were ``companion'' lightnings within the time window $\pm1$~s that
corresponds to the accuracy of the TUS trigger for the majority of slow
flashes. In most cases, the lightnings were registered at distances
$>400$~km from the center of the field of view of TUS.  It became clear
that the likely origin of slow flashes with uniform illumination of the
focal surface is light from distant lightning strikes that arrives at
the mirror due to the lack of any side shields and gets scattered.

\begin{figure}[!ht]
	\centering
	\mbox{\figh{.25}{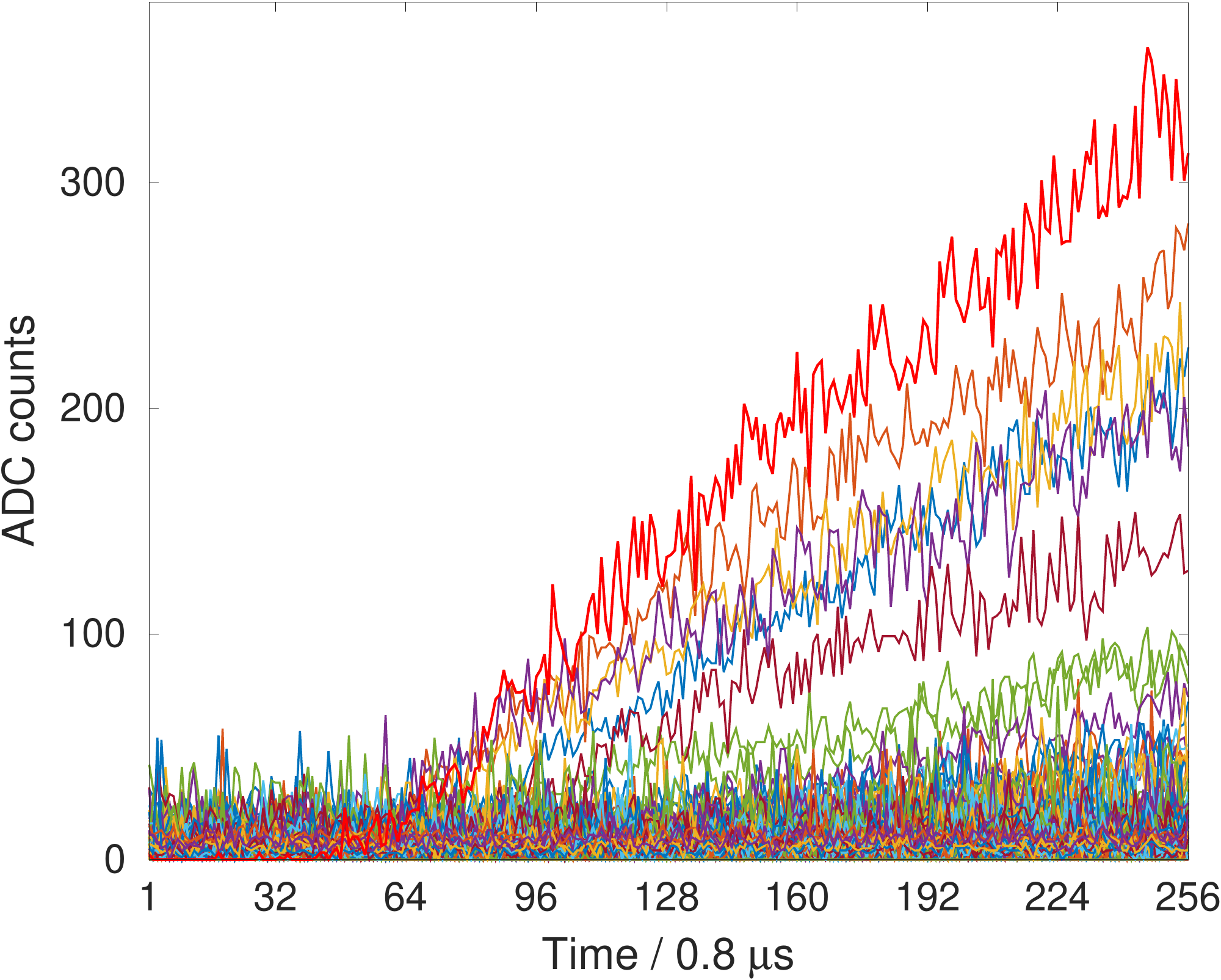}\quad\figh{.25}{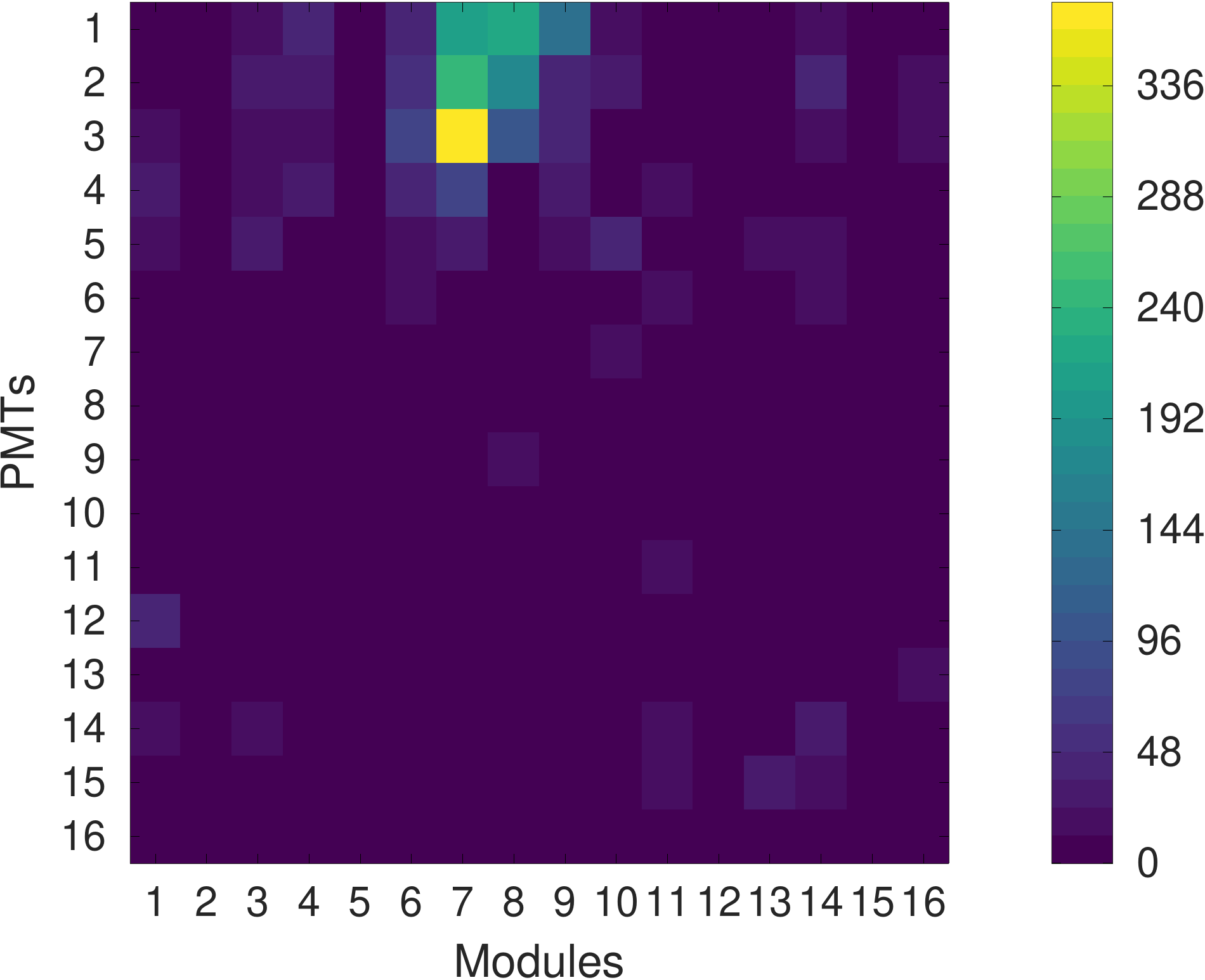}}
	\caption{Slow flash with strongly nonuniform illumination of the
	focal surface. Left: waveforms. Right: a snapshot of the focal
	surface at the moment of the maximum signal.}
	\label{sf-nonuniform}
\end{figure}

Illumination of the focal surface was found to be strongly non-uniform
in some cases while waveforms were typical for slow flashes. An example
is shown in Figure~\ref{sf-nonuniform}. In this particular case,
two lightning strikes were registered by
the Vaisala Global Lightning Dataset GLD360~\cite{vaisala1,vaisala2}
within $\pm1$~s from the TUS trigger time stamp approximately at the
position of the flash (within the FOV of TUS), see a bright spot in the
right panel of Figure~\ref{sf-nonuniform}.

\begin{figure}[!ht]
	\centering
	\fig{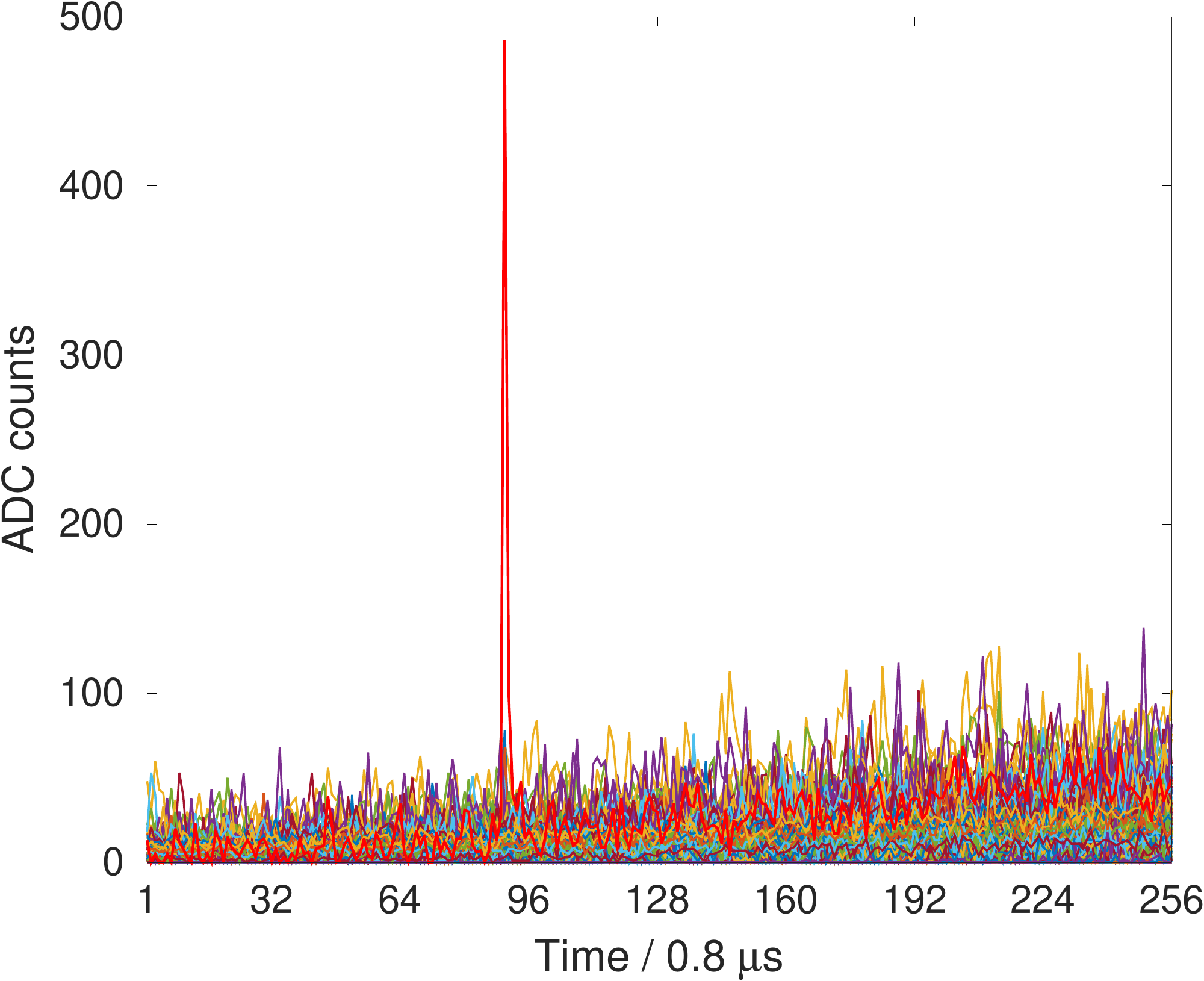}\quad\fig{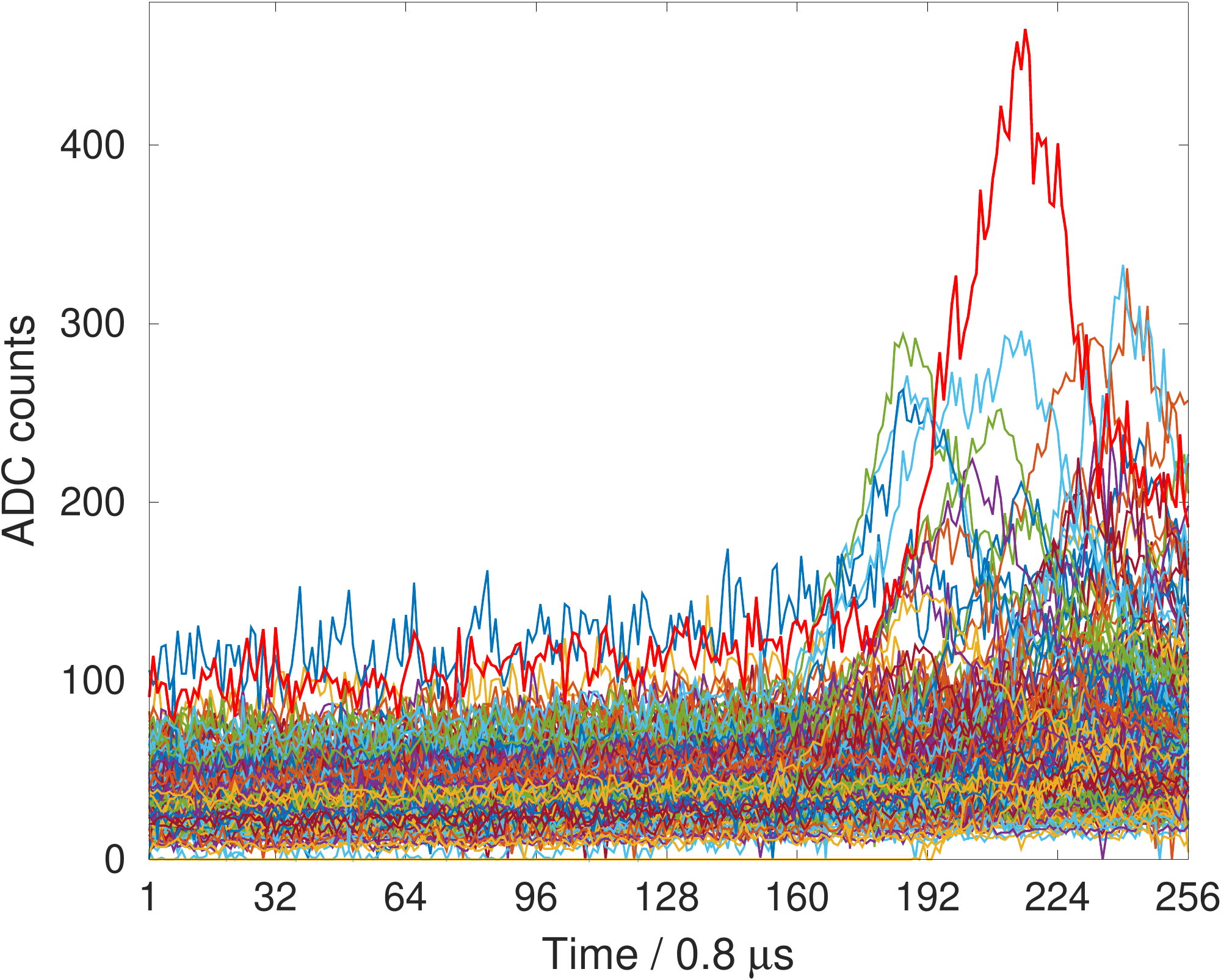}
	\caption{Slow flashes registered with other types of events.
		Left: a slow flash and a TLF. Right: a slow flash and an ELVE.}
	\label{sf-combined}
\end{figure}

Slow flashes were occasionally registered simultaneously with other
types of signals. The left panel of Figure~\ref{sf-combined} presents an
example of a slow flash registered together with an instant track-like
flash. The right panel of Figure~\ref{sf-combined} shows an ELVE
evolving simultaneously with a slow flash. Interestingly, at least 15
out of 26 ELVEs found in the TUS data demonstrated a similar behaviour.

Studies of slow flashes were continued after~\cite{JCAP2017}.
In particular, a data set of 3,327 events of this kind was obtained
with the help of a conventional algorithm that looked for an increase of
ADC codes by $\ge5\sigma$ over the background level in at least 20
channels~\cite{2020CosRe..58..317K}.

%______________________________________________________________________________
\section{Application of neural networks}

There are multiple approaches to classification of data based on machine
learning, see, e.g.,~\cite{Goodfellow}.  Here we consider two simple
architectures of neural networks.  This is justified by two reasons.
The first and most important one is that even these implementations
demonstrate a very high accuracy of classification.  Second, it is
interesting to estimate if a neural network can be used onboard a
satellite as a real-time trigger or anti-trigger for certain types of
events. One should employ a simple neural network in this case since
onboard processors have limited performance.

It is important to underline that TUS operated in different
observational conditions that used to vary due to changes of the phase
of the Moon, cloud coverage, humidity of the atmosphere, anthropogenic
lights etc. A natural way to mitigate these factors would be to
``flat-field'' signals in different channels, i.e., to adjust ADC codes
of all channels to a joint scale. Unfortunately, it is difficult to
implement this accurately because, as it was mentioned above, 51 PMTs
were totally damaged and sensitivities of the other channels changed
after a malfunctioning of the high voltage system during the first day
of TUS operation. A few attempts for an in-flight calibration of the
detector were performed but the estimates obtained are not accurate
enough for a reliable adjustment of ADC codes in different channels to a
joint scale.  This complicates a classification of the TUS data within a
uniform approach.

\subsection{Instant track-like flashes}

As it was already mentioned in the introduction, results of the first
application of neural networks to the analysis of the TUS data were
briefly reported in~\cite{BMY2020}. Let us recall the main
results of that study and discuss the subsequent analysis.

The main goal of~\cite{BMY2020} was to estimate if one can effectively employ
neural networks as triggers or anti-triggers for certain types of
signals that took place during the TUS mission and are likely to occur
in the future orbital fluorescence telescopes aimed at studies of
\uhecrs.  These should be simple neural networks since on-board
processors have limited performance.  The TLFs were chosen for the first
study because, on the one hand, they have a simple structure but on the
other hand they can hardly be avoided in orbital experiments.

Neural networks of two different architectures were considered.  These
were multilayer perceptrons (MLPs) and convolutional neural networks
(CNNs).  Perceptrons with 1, 2 and 3 hidden layers, consisting of 2--32
neurons were tested.  To keep things simple, input data for perceptrons
included only 512 numbers out of the whole set of 65,535 comprising a
TUS event. The first 256 numbers contained ADC codes of a waveform with
the highest amplitude and with the biggest FHDM in case there were
several waveforms with the same amplitude. The second half of the input
data represented a reshaped ``snapshot'' of the focal surface at the
moment of the first maximum. The aim of these numbers was to separate
TLFs from so called ``steep-front'' events that had waveforms of the
similar shape but used to illuminate the whole focal surface.

Results presented in~\cite{BMY2020} were mainly based on employing data
of two months of observations with a subset of previously marked TLFs as
a training data set. The data set included results of observations
during October, 2016, and April, 2017. The choice was justified by a few
considerations. October, 2016, was approximately in the middle of the
first period of regular observations in the EAS mode that lasted from
August~16 till December~16, 2016. It contained 7,888 events including
961 previously known TLFs.  (Twenty four other TLFs were found due to
using NNs.)  The second month contained data different from the first
one in two respects. First, a preliminary version of an anti-trigger
aimed to suppress TLFs was implemented in the onboard software of TUS in
late March, 2017. It did not work effectively enough but resulted in a
shift of the moment of a signal rise in TUS records from time step
around 60 to that around 30. This could influence the accuracy of neural
networks.  Second, another anti-trigger was uploaded to TUS on April~28,
2017.  This anti-trigger effectively suppressed the majority of TLFs but
left space for ``tails'' of the most long-lasting of them, see an
example below.  Such events were necessary for a proper training of the
NNs.  The set recorded in April, 2017, consisted of 8,103 events,
including 370 TLFs.  We tried several other ways of selecting training
data sets within the present work, including random sampling, but they
did not demonstrate any considerable benefits in comparison with the
initial choice.

The main part of the analysis presented in~\cite{BMY2020} employed MLPs
with two hidden layers with 8 neurons in each of them.  The TLFs were
selected in advance with a conventional algorithm that looked for an
instant growth of the signal over the background by more
than~$20\sigma$.  Such perceptrons demonstrated an accuracy $\ge0.995$
on the training data but the accuracy of even more simple neural
networks was not drastically lower.  The accuracy of classification on
the test data sample was also pretty high: the networks used to make
$\sim200$ wrong predictions in the data set of more than 61 thousand
events.  The majority of mistakes related to low-amplitude flashes
confined to a single pixel with a comparatively high level of the
background signal similar to the one shown in the right panel of
Figure~\ref{tracks-waveforms}.  This kind of signals was also the most
difficult one for the conventional algorithm.

We also tested a number of configurations of CNNs. The network used
in~\cite{BMY2020} consisted of a convolutional layer with 20 filters and a
$3\times3$ kernel, a maximum pooling layer, a hidden dense layer with 64
neurons and an output layer. In the subsequent analysis reported below,
the CNN was complemented with a dropout layer and a dense hidden layer
with 256 neurons going after the pooling layer. The output layer used
sigmoid as an activation function; the other layers used ReLU. The Adam
algorithm was used as an optimizer. A procedure for an early stop of
training to prevent an overfitting of the neural network and an
adjustable learning rate available in Keras were employed.

The highest accuracy of the CNN was obtained with every event in the
input sample represented by 768 numbers. These were ``snapshots'' of the
focal surface made at the first moment of an event, at the moment of the
first maximum of ADC codes and at the last moment. These snapshots were
arranged in three layers so that the signal in each pixel was described
by three numbers similar to RGB images.  Table~\ref{table1}
provides some numbers that illustrate the training stage of the CNN and
a number of MLPs with two hidden layers but different number of nodes in
each of them (from~4 to~48). The sample consisted of nearly 16,000
events, two thousand of which were used for testing the trained
networks.  Twenty percent of the rest of the sample were employed as a
validating sample.  Neural networks of each configuration were trained
ten times with different initial weights.  It can be seen that
performance of MLPs was very high in all cases and only slightly
depended on the number of neurons in the hidden layers. The CNN
demonstrated higher accuracy than MLPs during training but its testing
accuracy was slightly lower than for MLPs with more than~24 neurons in
each hidden layer. We do not include data for perceptrons with more
hidden layers or neurons because we did not find an increase of
accuracy with such networks.

\begin{table}[ht]
	\centering
	\caption{Mean accuracies observed during the training stage of the CNN
	(the last column) and 2-layer MLPs with different number of nodes
	in each of the hidden layers (4, 8,\dots, 48).}
	\medskip
	\begin{tabular}{|l|c|c|c|c|c|c||c|}
		\hline
		2-layer MLPs. \# nodes:	& 4 & 8 & 16 & 24 & 32 & 48 & CNN \\
		\hline
		Training accuracy       &0.9983&0.9991&0.9993&0.9995&0.9996&0.9997&1.0\\
		\hline
		Best validation accuracy&0.9954&0.9960&0.9969&0.9966&0.9969&0.9972&0.9976\\
		\hline
		Testing accuracy  		&0.9954&0.9961&0.9969&0.9973&0.9974&0.9975&0.9973\\
		\hline
	\end{tabular}
	\label{table1}
\end{table}

\begin{figure}[!ht]
	\centering
	\fig{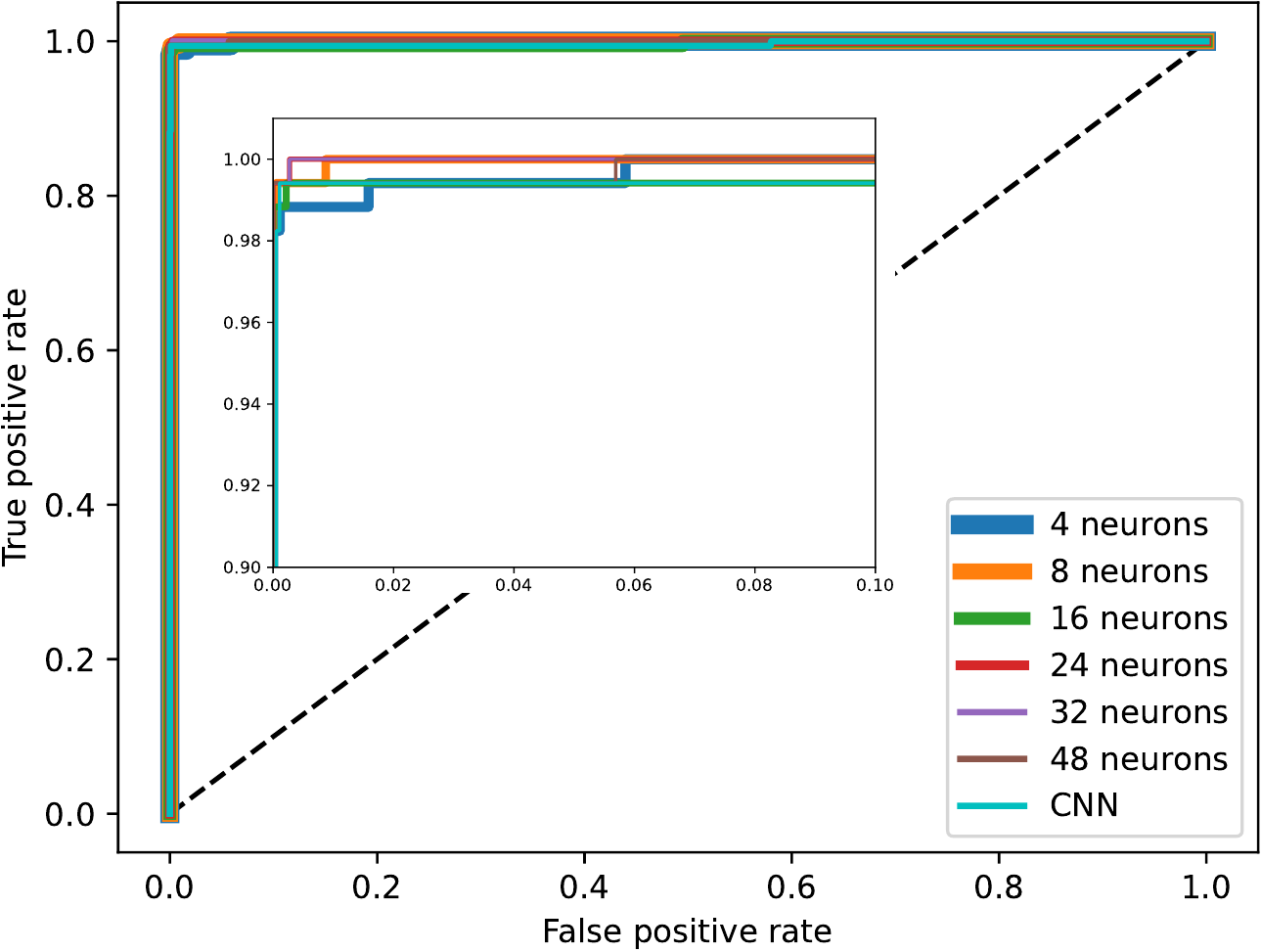}\quad\fig{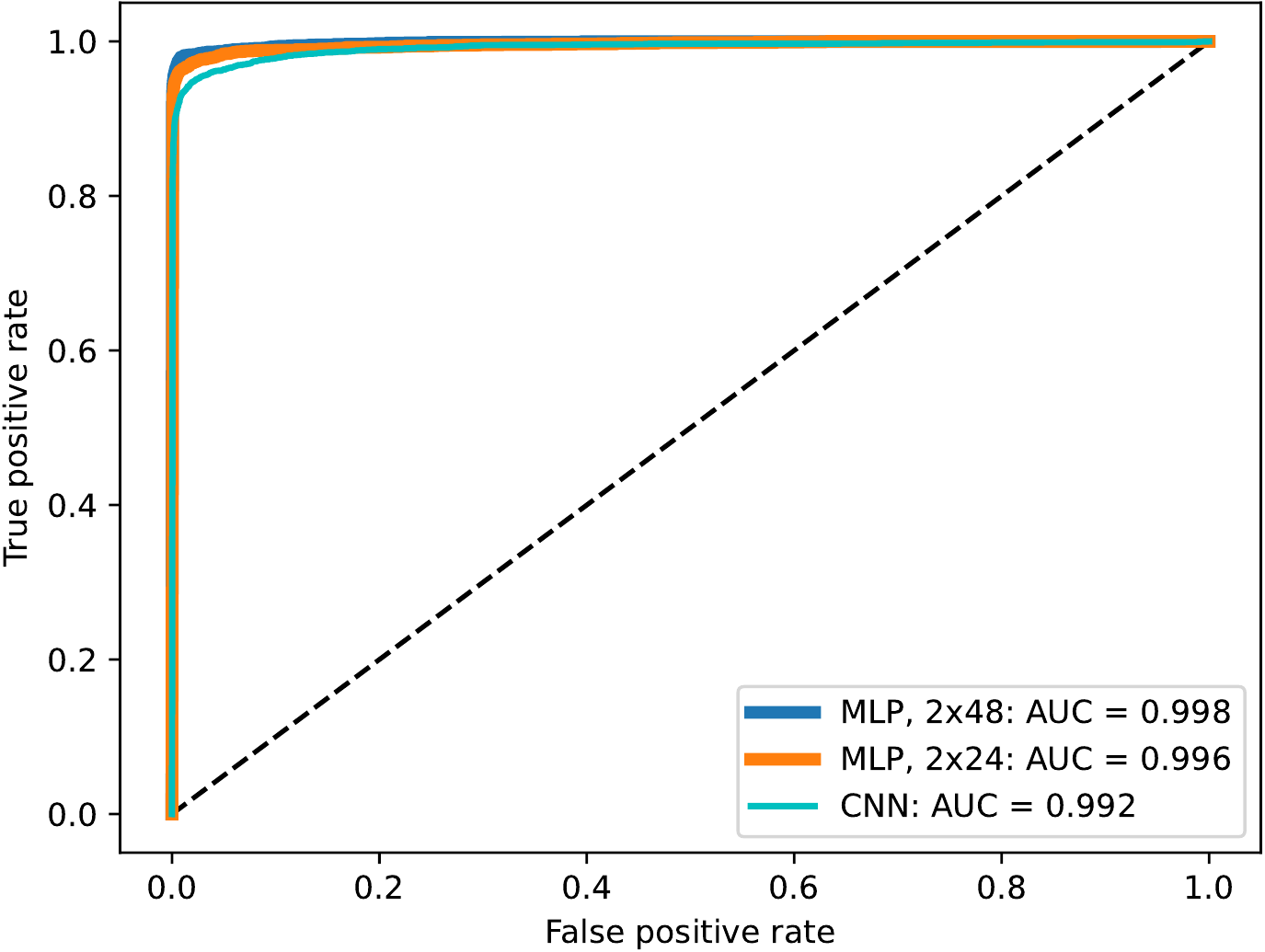}
	\caption{ROC curves for the training (left) and testing (right)
	stages for the CNN and MLPs with 2 hidden layers and different number
	of nodes in each layer.}
	\label{tlf-roc}
\end{figure}

Figure~\ref{tlf-roc} presents ROC (Receiver Operating Characteristic)
curves that illustrate the efficiency of different architectures of
neural networks employed in searching for TLFs.  The left panel shows
ROC curves for the training stage of the CNN and 2-layer MLPs with 4,
8,\dots, 48 neurons in each hidden layer.  It is clearly seen that there
is very little difference in performance of MLPs with $\ge8$ nodes and
the CNN. The right panel shows ROC curves for the testing stage of the
CNN and 2-layer MLPs with 24 and 48 nodes in each layer.  In this case
the MLPs performed slightly better than the CNN with the area under the
curve (AUC) equal to 0.996 and 0.998 respectively in comparison with
0.992 for the CNN.  A reason why MLPs slightly outperformed the CNN
might be the specific type of waveforms in track-like flashes that can
be easily recognized by an MLP (in most cases) and a much more rich
variety of their traces on the focal surface, partially due to dead or
buggy pixels and some glitches of the electronics discussed above.

%The average accuracy of the CNN
%was slightly lower than of the 2-layer perceptron, with typically around
%350 mistaken classifications in the testing sample. 

\begin{figure}[!ht]
	\centering
	\fig{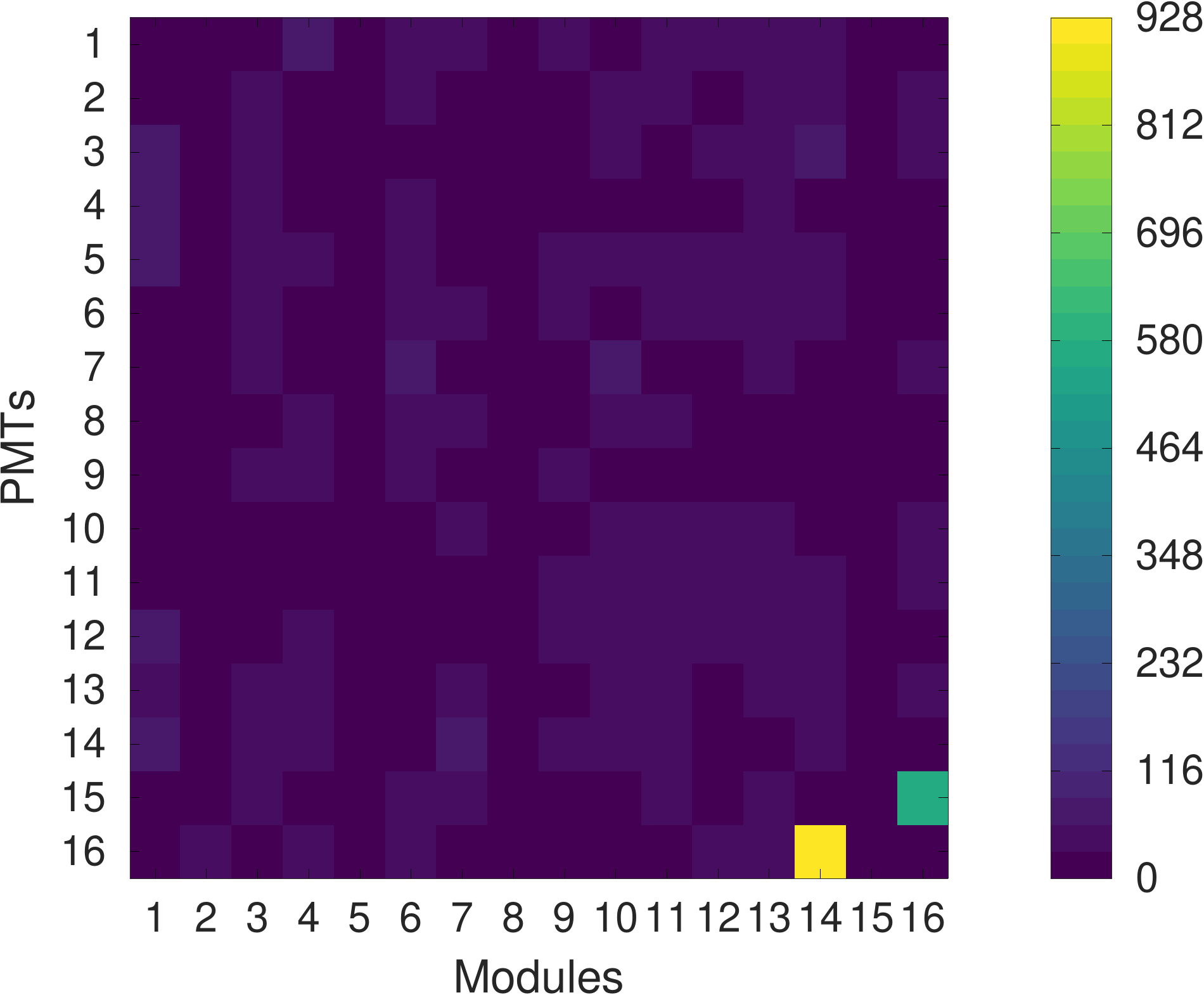}\quad\fig{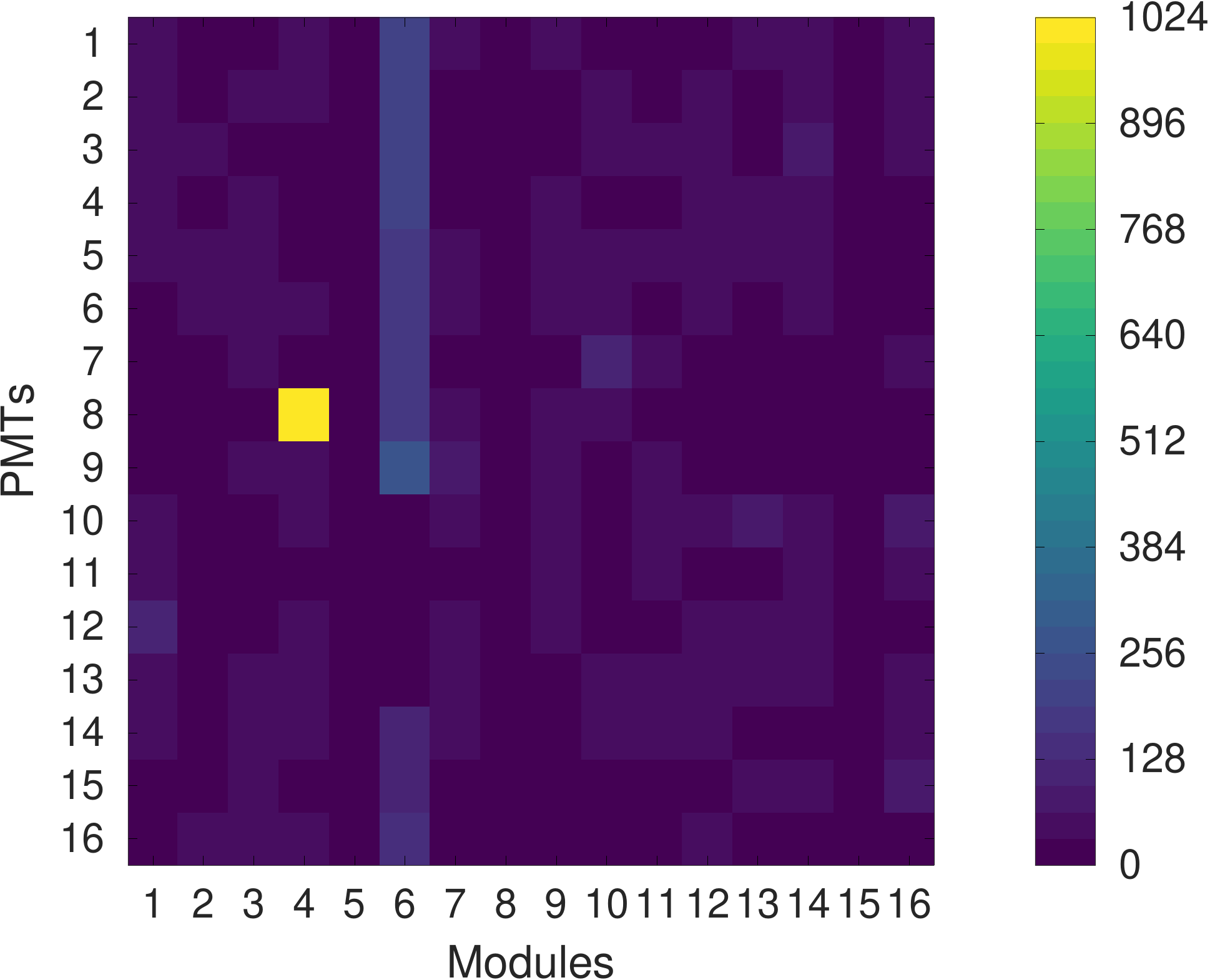}\\[2mm]
	\fig{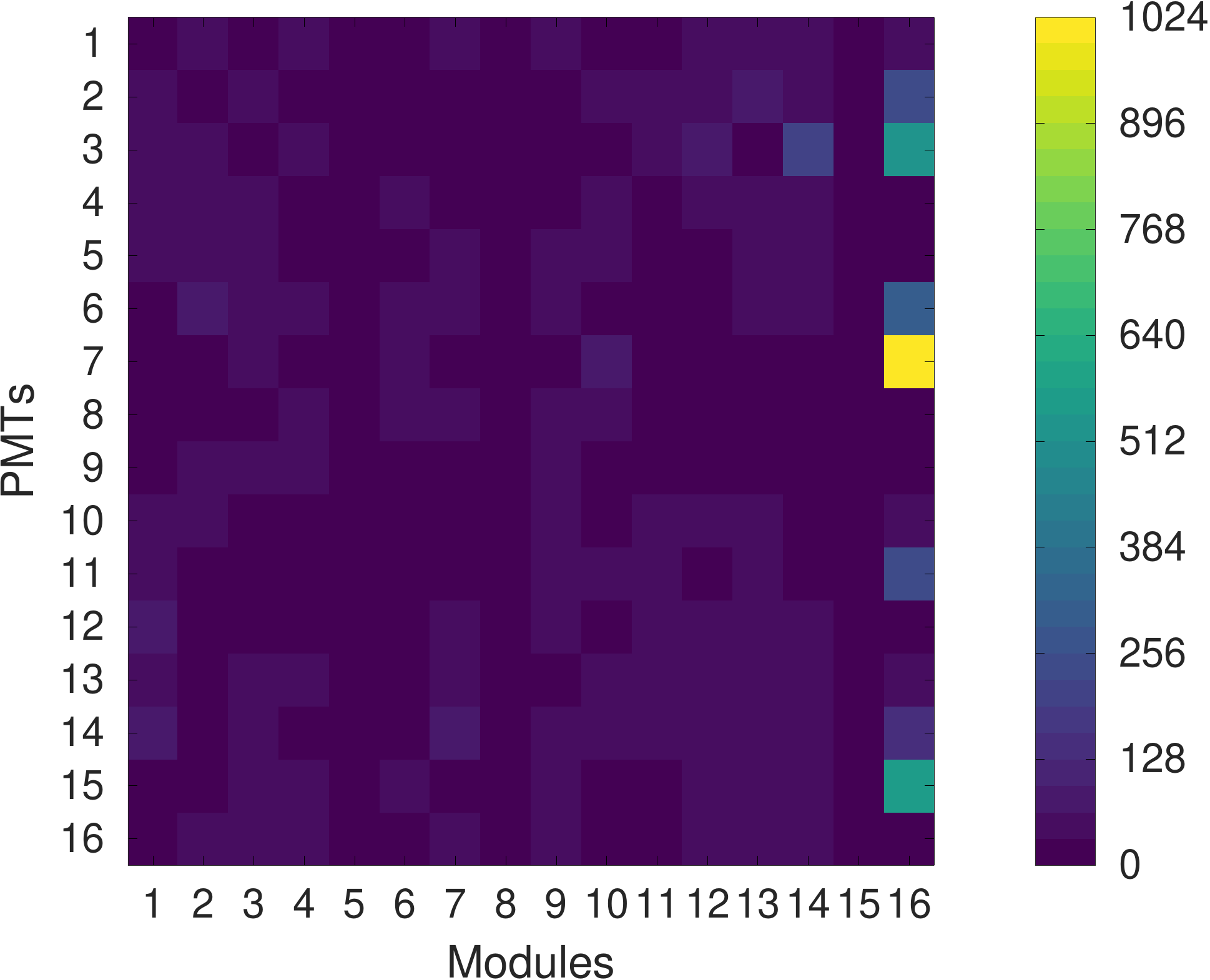}\quad\fig{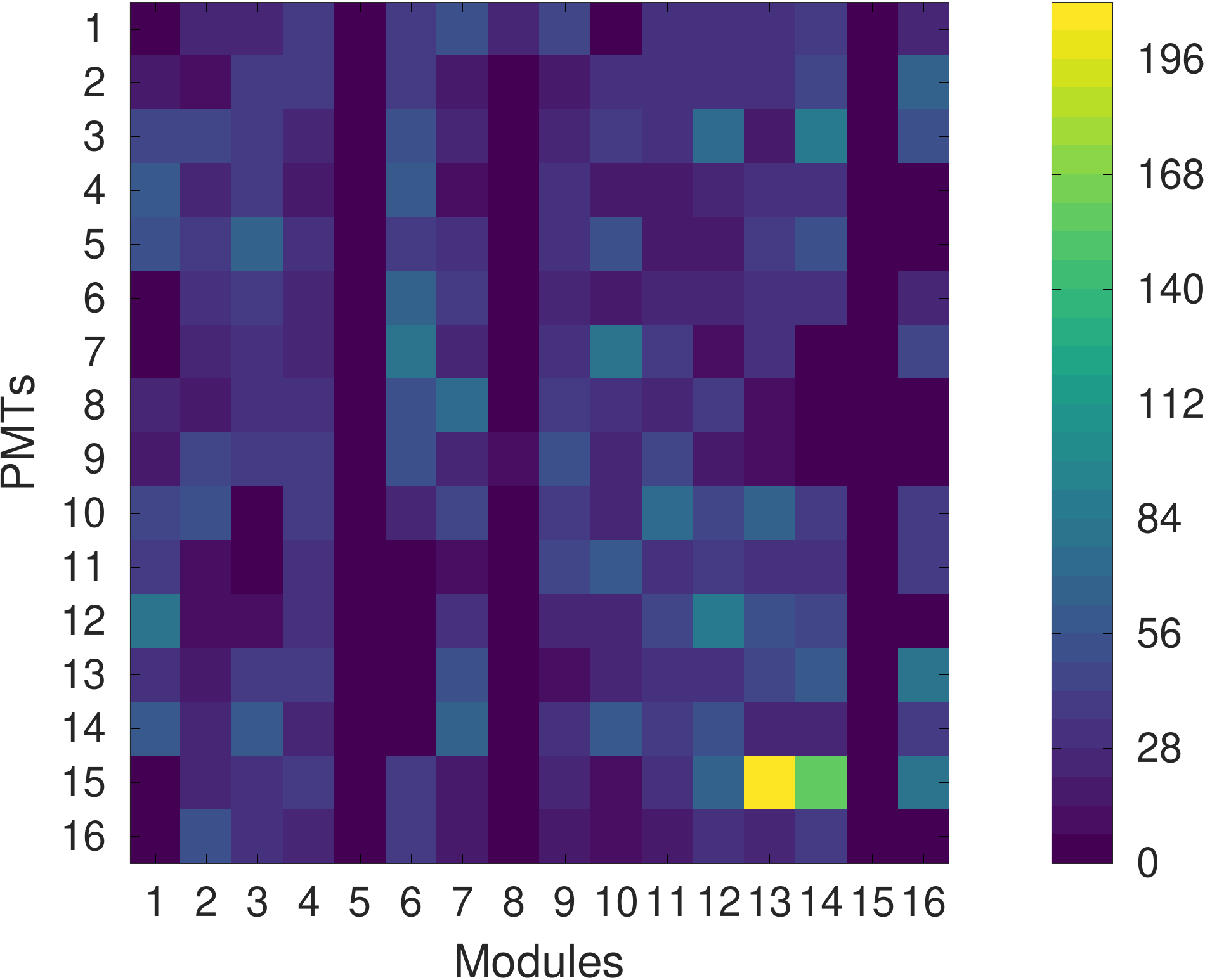}
	\caption{Examples of snapshots of the focal surface that were
	not classified by the CNN as produced by TLFs.}
	\label{tracks-lost}
\end{figure}

\begin{figure}[!ht]
	\centering
	\mbox{\figh{.25}{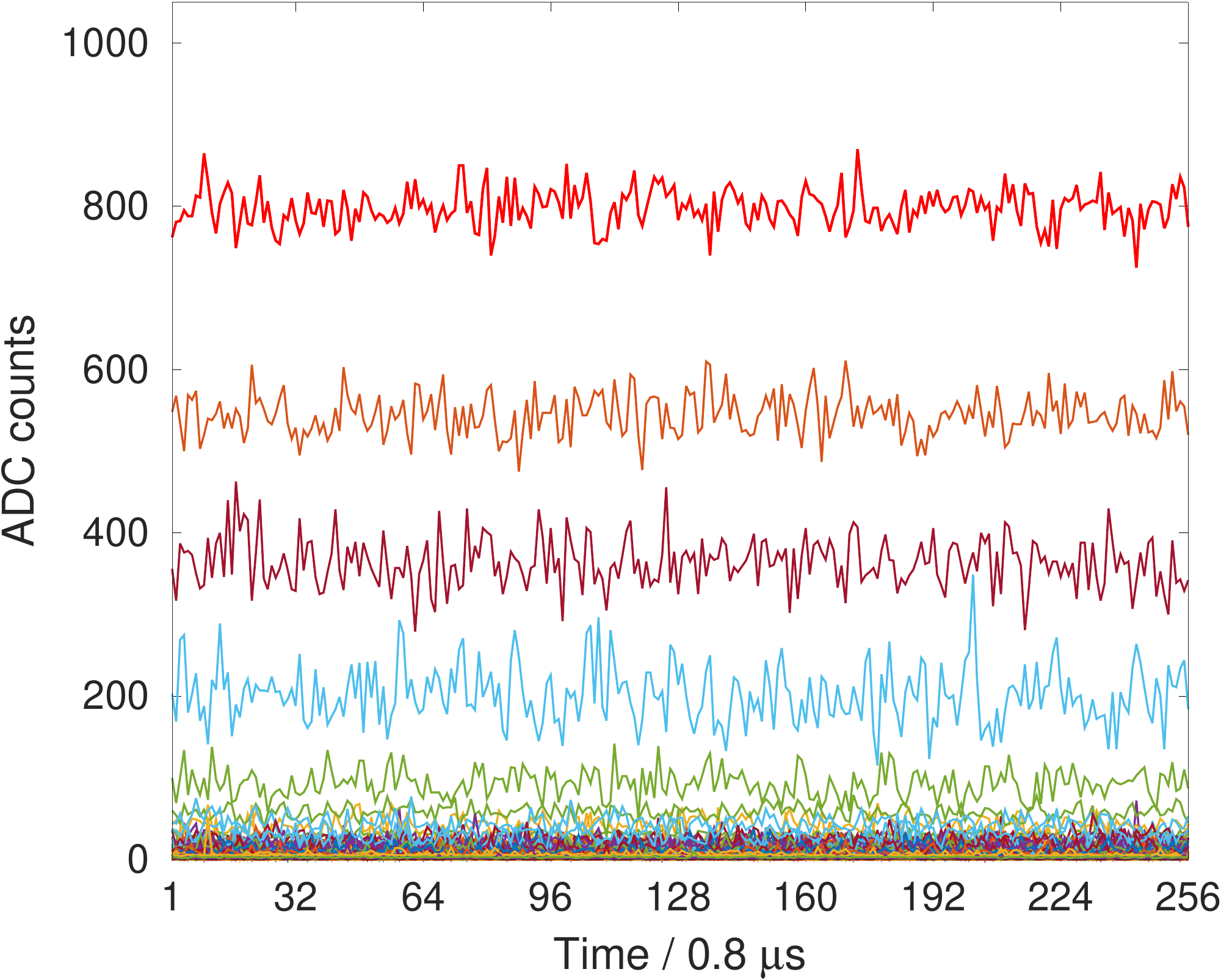}\quad\figh{.25}{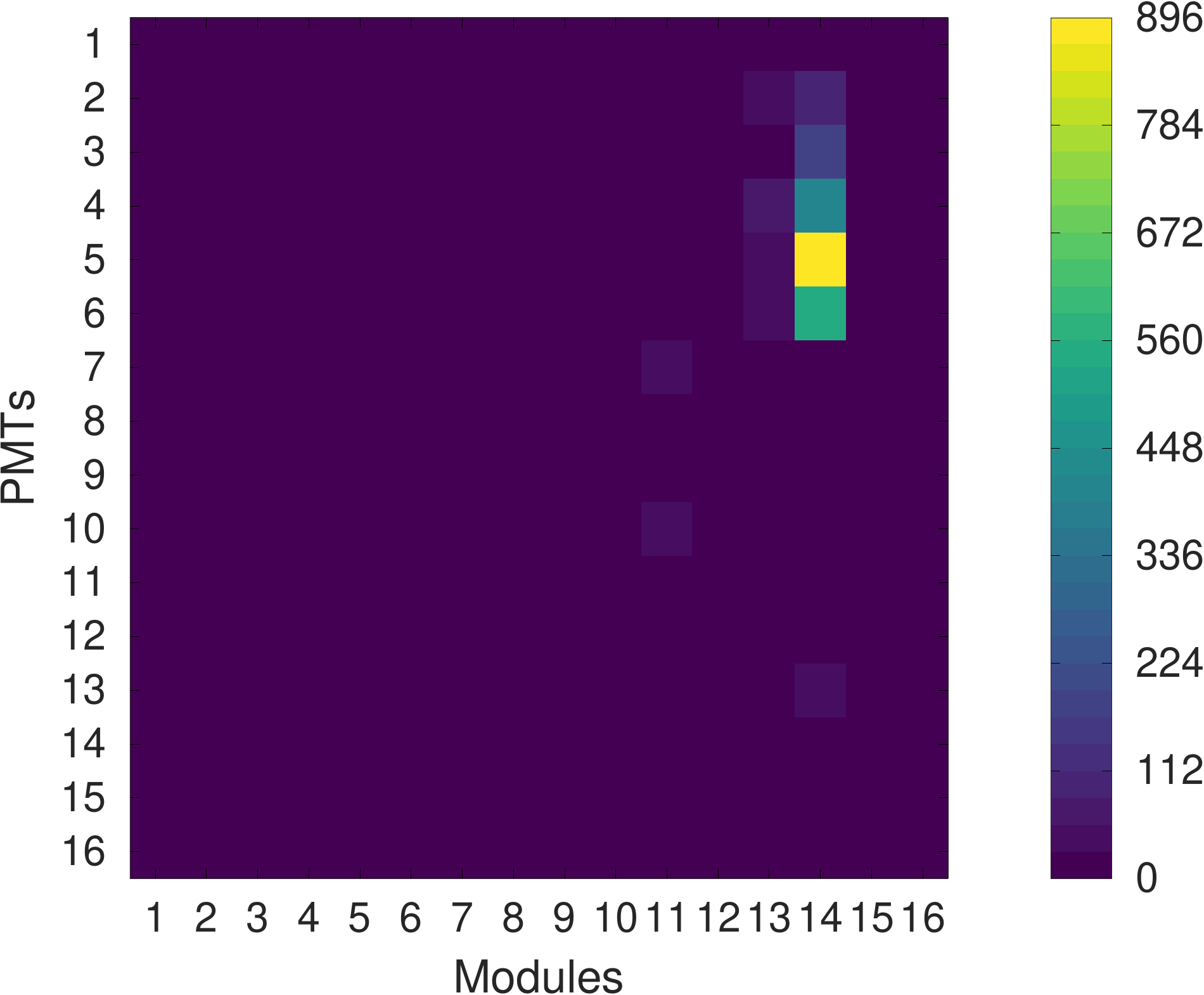}}\\[2mm]
	\mbox{\figh{.25}{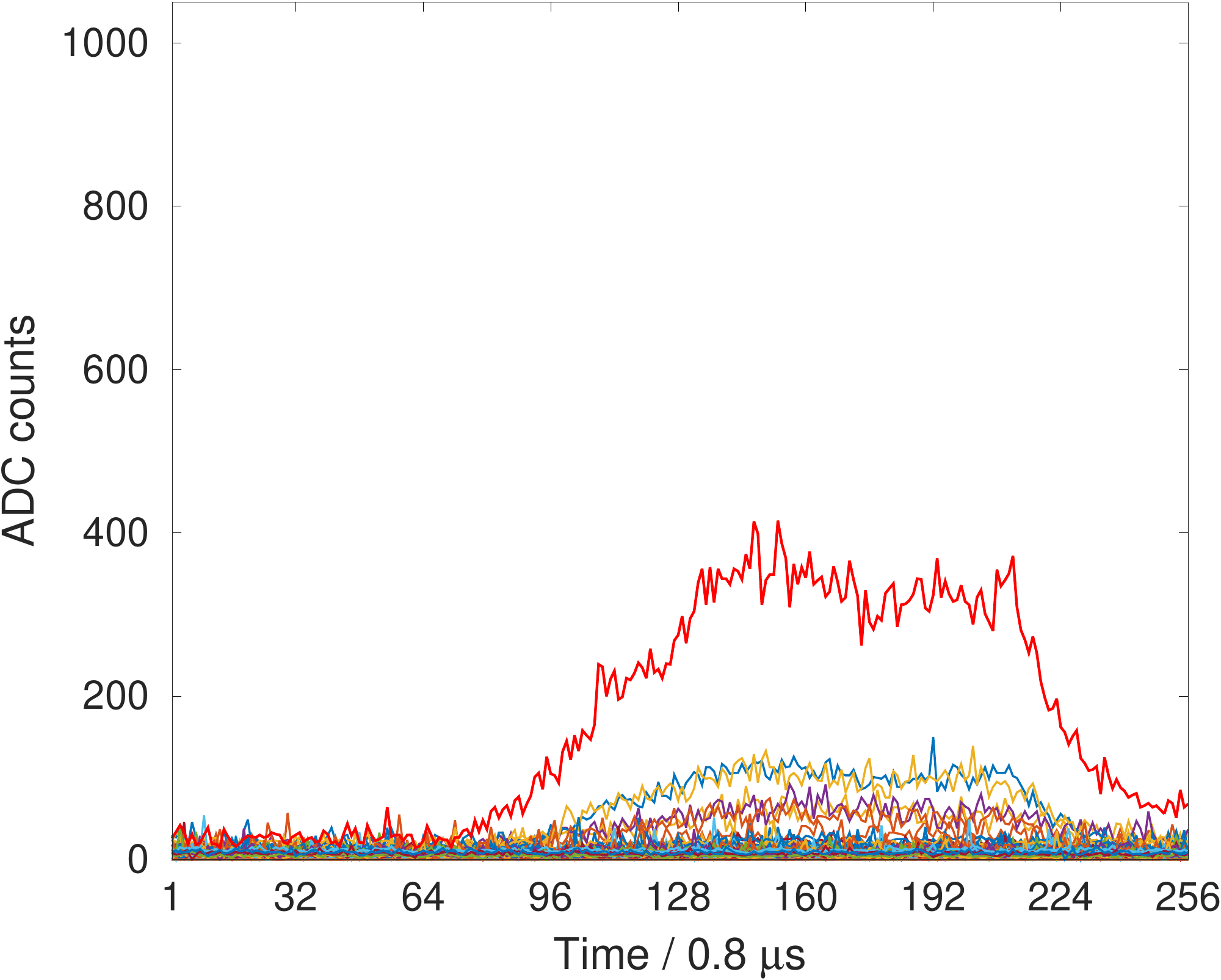}\quad\figh{.25}{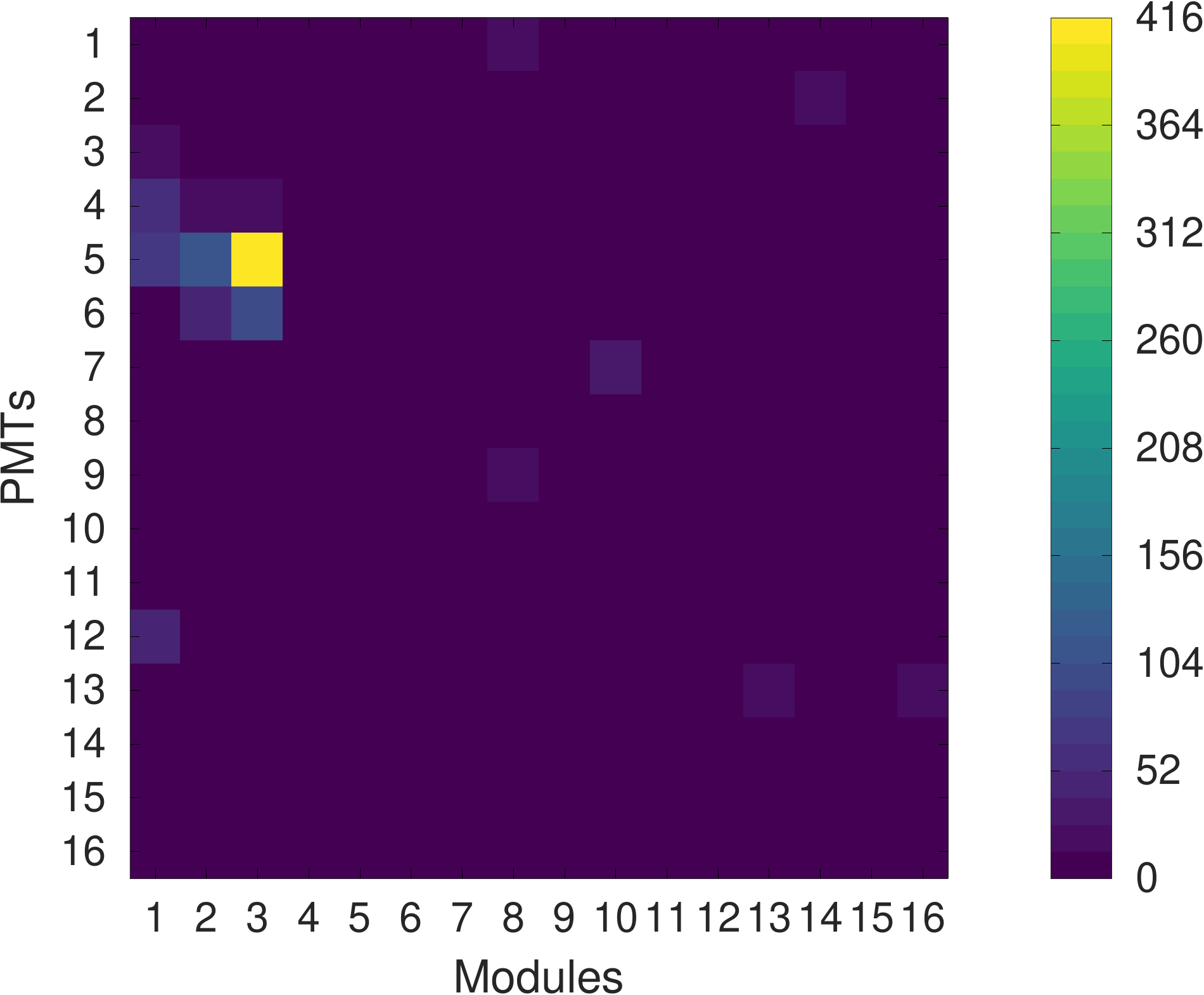}}
	\caption{Examples of events that were mistakenly classified by
	the CNN as instant TLFs.}
	\label{tracks-false}
\end{figure}

A few examples of snapshots of the focal surface of TLFs that were not
properly classified by the CNN are shown in Figure~\ref{tracks-lost}. On
the other hand, there were a number of events that were mistakenly
classified as TLFs by the CNN.  Two examples of such false positives are
presented in Figure~\ref{tracks-false}.  It is immediately apparent from
the waveforms that the events were not track-like flashes but the
snapshots were confusing for the neural network.  Events with tracks
similar to those shown in Figure~\ref{tracks-electro} also posed a
problem for the CNN.  It is unlikely that issues like these will take
place with orbital fluorescence telescopes with higher spatial
resolution, smaller pixels and other electronics systems.

\begin{figure}[!ht]
	\centering
	\fig{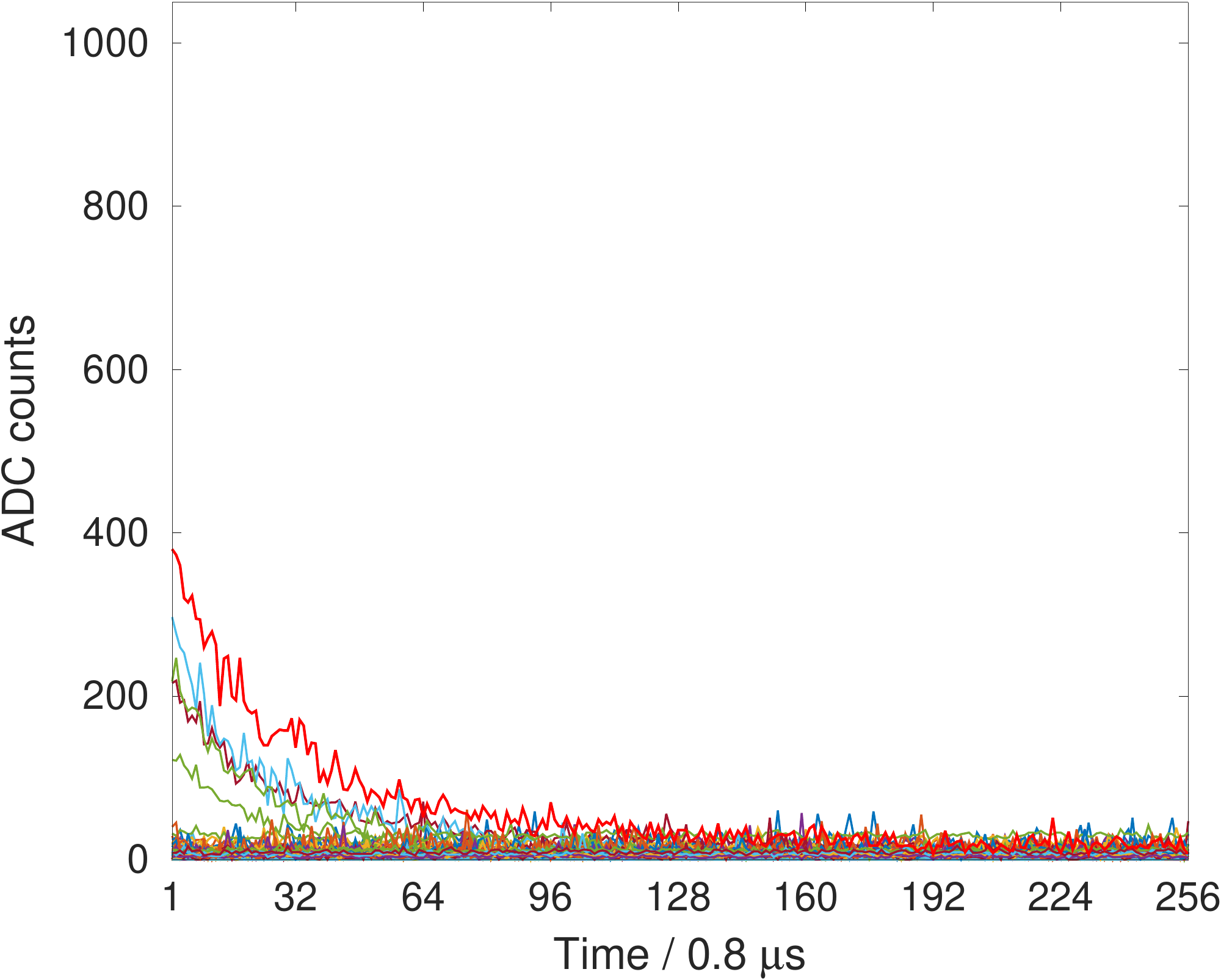}\quad\fig{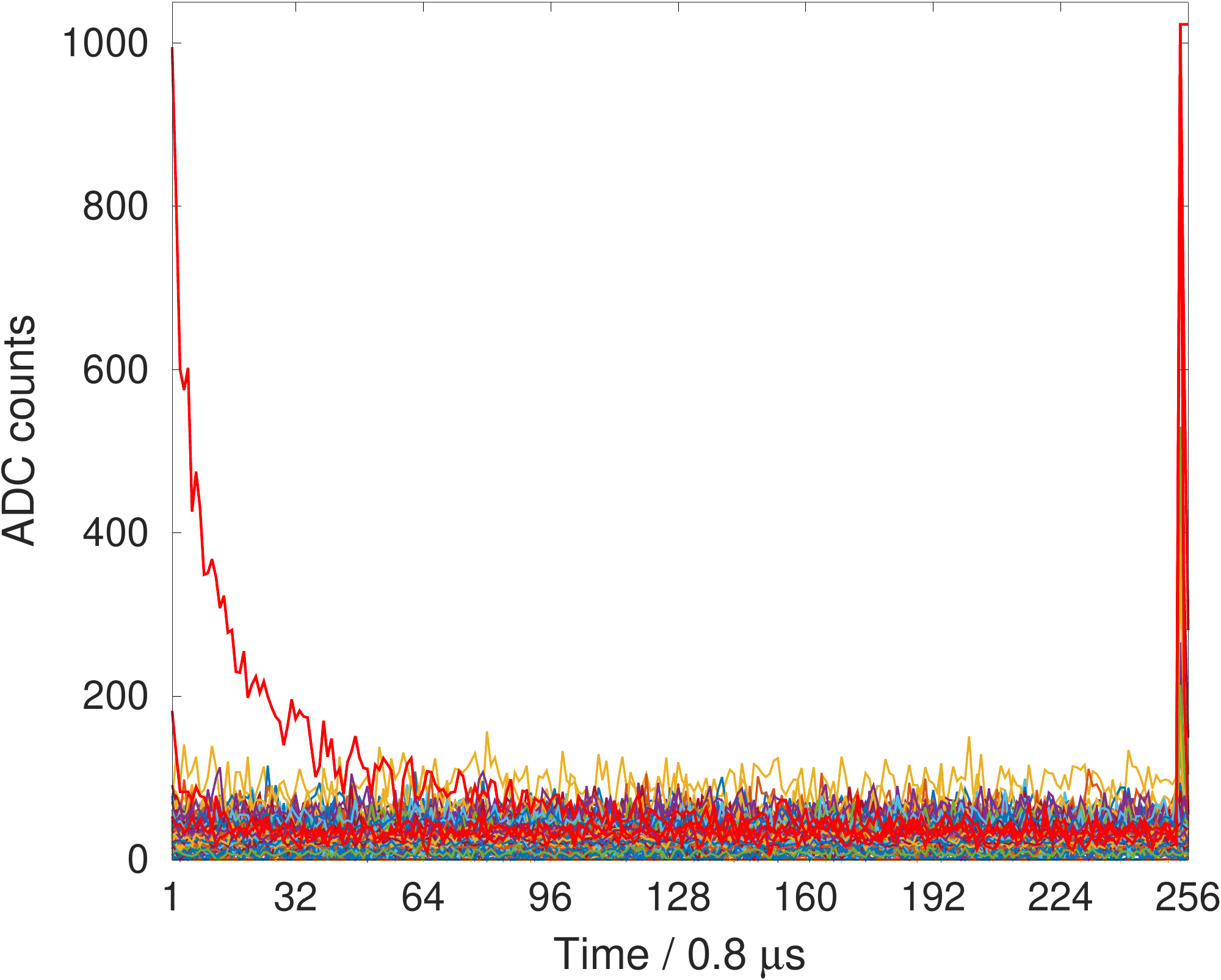}
	\caption{Peculiar waveforms of some TLFs. Left: a ``tail'' of a TLF
	suppressed by the anti-trigger introduced in late April, 2017.
	Right: waveforms of a TLF with a mistakenly placed beginning of the
	record due a bug in the onboard software.}
	\label{tracks-cnn}
\end{figure}

Neither of the two false positives shown in Figure~\ref{tracks-false}
(nor similar ones) posed a problem for the MLPs.  However, the CNN was
more effective when detecting ``tails'' of TLFs that used to occur from
time to time in the data after the majority of TLFs were suppressed by
the anti-trigger, see the left panel in Figure~\ref{tracks-cnn} for an
example of such event.  More than this, the CNN trained on data
represented by 768 numbers in the way described above was able to
classify properly TLFs like the one shown in the right panel of
Figure~\ref{tracks-cnn}. Such events used to appear in the period from
May till August 2016 due to a bug in the onboard software that resulted
in an incorrect sequence of ADC codes, putting the first moment of time
somewhere around the middle of the record.

Interestingly, neither increasing the amount of data used for a
description of every TUS event, nor a few modifications of the CNN lead
to an improvement of the accuracy of classification.  Even CNNs
employing all available ADC codes in events used for training
demonstrated lower accuracy on the testing sample than the CNN employing
768 numbers.  They were sometimes confused by events in which the peak
of the signal was located close to the beginning or to the end of the
record.  This might by due to the fact that the moment of the instant
growth of the signal was not always the same in the TUS data.  On the
one hand, CNNs that employed complete records for training resulted in
classifying very weak signals on a strong background as TLFs which we
were unable to verify as being such.

Remarkably, the neural networks found more than 400 weak TLFs not
identified as such by the conventional algorithm.
This increased the number of TLFs found in the TUS data by approximately
10\% resulting in nearly 4,000 events in the whole TUS data set.
(Obviously, one could lower the threshold of the signal jump in the
conventional algorithm but this used to lead to a high number of
false positives caused by kinks that were likely to be just random
fluctuations of the signal.)

One can notice that the training data set used in~\cite{BMY2020} was
strongly imbalanced with TLFs comprising only around 8.5\% of the whole
training sample.  It is known that using imbalanced data can lead to
a sub-optimal accuracy of a neural network.
We studied this issue by creating random samples of data used for training
instead of using continuous periods of time, with TLFs representing from
20\% to 50\% of the whole samples. We did not find any significant
improvement of the accuracy of NNs trained on these samples when applied
to testing data.  More than this, in some cases this resulted in a
growth of the number of false positives.

\begin{figure}[!ht]
	\centering
	\figw{.9}{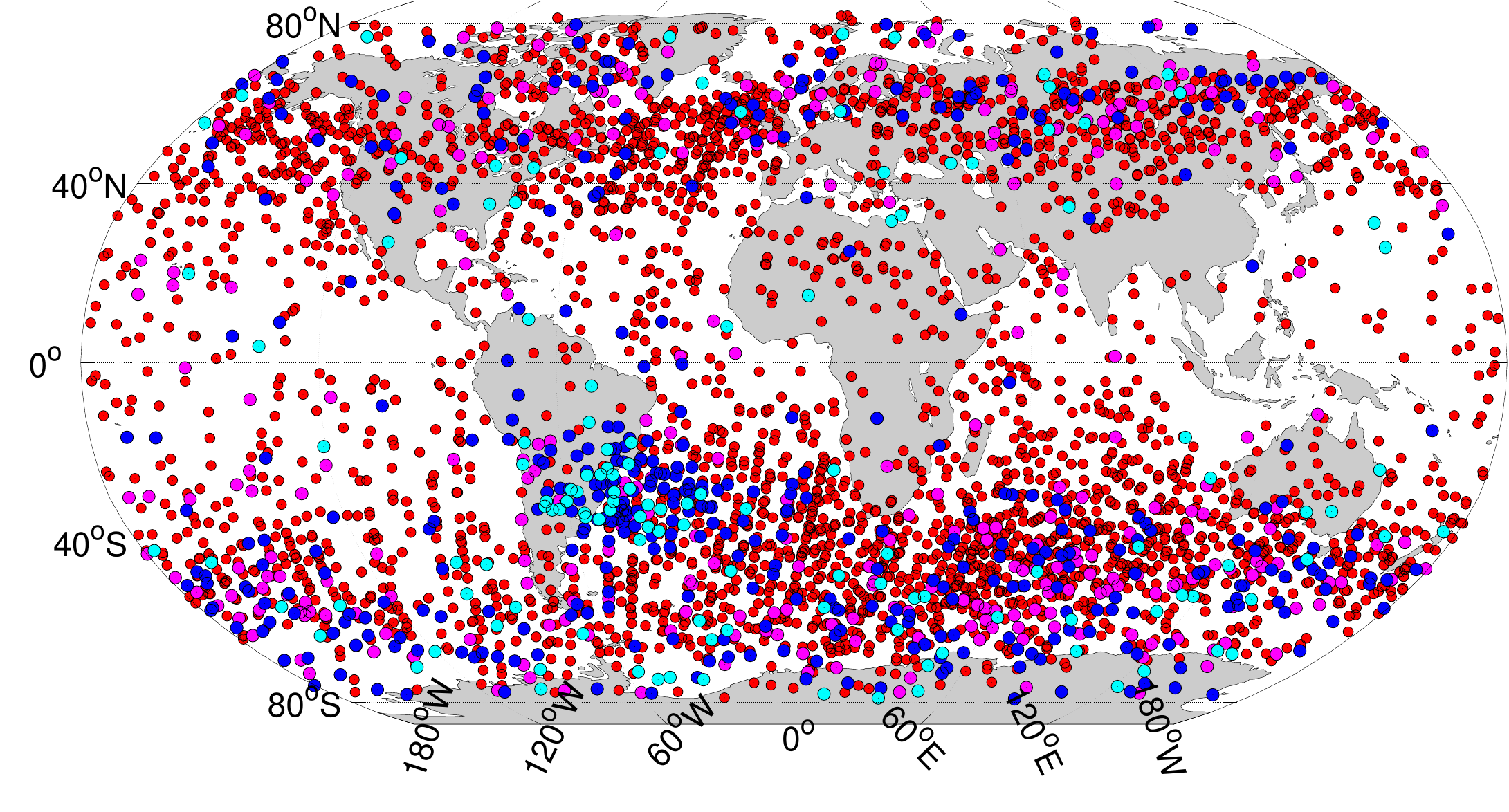}
	\caption{Geographical distribution of instant track-like flashes.
		Colours denote ranges of amplitudes: cyan, blue and magenta mark
		amplitudes $\le200, 400, 800$ respectively. Flashes with
		amplitudes $>800$ are shown in red.}
	\label{tlf-map}
\end{figure}

Geographical locations of instant track-like flashes in the final sample
are shown in Figure~\ref{tlf-map}. It is clearly seen that the majority
of TLFs had high amplitudes. In fact, 2,787 of them had amplitudes equal
to 1023 ADC counts thus saturating PMTs. Only 131 TLFs in the sample had
amplitudes $\le200$. The majority of them were found in the region of
the South-Atlantic anomaly, a huge area in the Southern hemisphere,
where the inner Van Allen radiation belt comes closest to the Earth's
surface, going down to an altitude of around 200~km, which leads to
an increased flux of energetic particles in this region.%
\footnote{\url{https://heasarc.gsfc.nasa.gov/docs/rosat/gallery/misc_saad.html}}

%______________________________________________________________________________
%\newpage
\subsection{Slow flashes}

Every TUS event can be considered as a video consisting of 256 frames
with a resolution of $16\times16$ each. One of the established
approaches for classification of videos is using so called convolutional
long short-term memory neural networks, see, e.g.,~\cite{ConvLSTM}.
However, we found them more demanding on computational resources than
the CNN considered here or its slight modifications without any
noticeable benefit in terms of accuracy.  Thus, in what follows, we only
consider an application of the CNN described above with different ways
of arranging input data.

It was clear from the very beginning that snapshots of TUS records made
at the first and last time steps and at the moment of the signal maximum
will be insufficient for an effective representation of slow flashes
because they do not have a characteristic instant ``jump'' of the
signal as TLFs. We tried several ways of preparing input data, mainly by
using samples uniformly distributed within a record.  We tested input
data represented as 4, 8, 16, 32, 64, 128 and 256 snapshots of the focal
surface arranged in a stack (a ``sandwich'') with the respective number
of layers.
None of the different ways of arranging input data demonstrated
significant benefits over the other in terms of the accuracy. In a
typical run, the best testing accuracy was approximately equal to 0.999,
the best validation accuracy was of the order of 0.99, and the testing
accuracy varied in the range 0.980--0.988. However, the time needed for
training used to grow considerably with an increasing number of layers.
We have chosen 16 snapshots of the focal surface uniformly
distributed within a record as the main way of representing input data.
Thus, only 4096 numbers (ADC codes) of the total 65,536 were used to
represent each of the TUS events.

We used different samples for training the network. In what
follows, we discuss results obtained by employing a sample of 16
thousand events registered during October, 2016, and July, 2017.
One thousand events acted as a testing sample, others were used for
training with 20\% of them forming a validation set.

\begin{figure}[!ht]
	\centering
	\fig{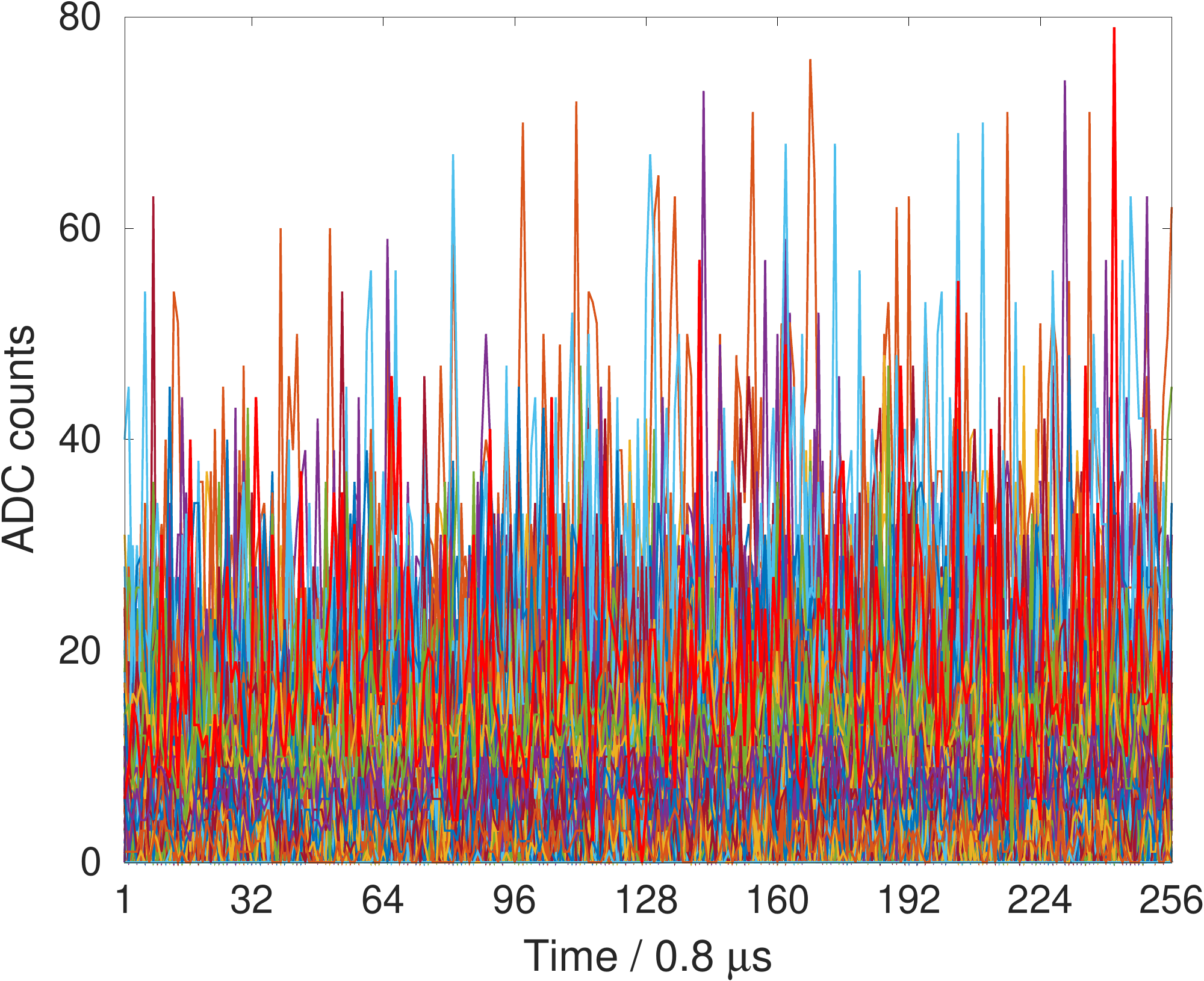}\quad\fig{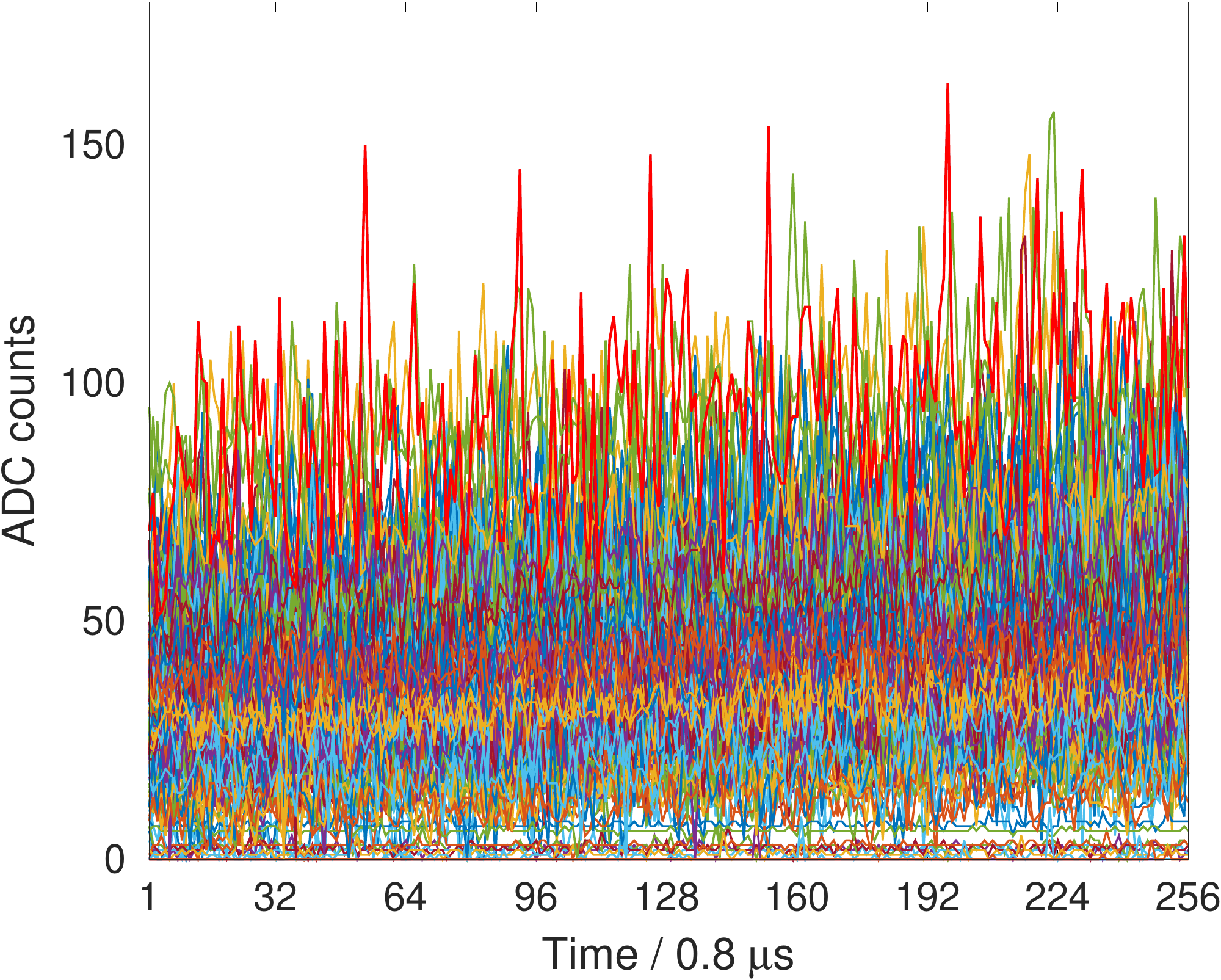}
	\caption{Two examples of events selected by the CNN as slow flashes
		but not know as such previously.}
	\label{sf-weak_new}
\end{figure}

The process of training the CNN brought a number of surprises.
Initially, the training sample included 798 slow flashes selected by a
conventional algorithm.  Training the network resulted in a considerable
number of ``false positives'' that either had low amplitudes without
obvious signs of a ``collective'' behaviour similar to that shown in
Figure~\ref{sf-typical}, or were registered on a strong background,
mostly during periods around a full Moon.  A closer inspection of these
events revealed that they could possibly be slow flashes. A couple of
examples of such events is presented in Figure~\ref{sf-weak_new}.

At this stage, we focused on a data set registered from August~16, 2016,
till September~19, 2017, since we had the WWLLN data for it.  The
initial sample of slow flashes for this period consisted of 175 events
(out of 6,584 events in the whole sample).  An analysis of all ``false
positives'' selected by the CNN revealed a correlation between all new
events selected around the full Moon and 76 out of 78 weak events on the
one hand and companion lightnings/thunderstorms on the other hand.  In
particular, 5 strikes were registered by the WWLLN in less than 1~s
after the event shown in the left panel of Figure~\ref{sf-weak_new} in
440--470~km from the center of the FOV of TUS. Another strong
thunderstorm was taking place at the moment of registering an event
shown in the right panel of Figure~\ref{sf-weak_new} with~6 lightnings
registered within 1~s from the TUS trigger time stamp at distances
around 680~km. This could not unequivocally prove that these events were
due to distant lightning strikes but provided an evidence that the CNN
might be more sensitive to weak slow flashes than the ``traditional''
algorithm and that at least a part of events selected as ``false
positives'' could be slow flashes.

Since we did not have data on lightning strikes for the rest of the TUS
data, we looked for another conventional tool that could help in judging
if an event newly selected by the CNN could be a slow flash.  We have
found that waveforms of almost all slow flashes (except the most violent
ones that resulted in a saturation of the focal surface) could be nicely
smoothed with a Hamming window of size 128 and fitted with a Gaussian,
and the number of good fits was $\ge10$.  Similar to correlations with
lightnings, this test could not prove that a particular event was or was
not a slow flash in terms of its origin related to lightning strikes.
However, we employed it in later studies to be on the conservative side
when checking doubtful events selected by the CNN.

To deal with the incompletely marked training data set, we applied the
following iterative procedure.  After training the CNN, we visually
checked all new ``false positives'' and in case of doubts applied the
above mentioned ``smooth-and-fit'' procedure to them. Then we repeated
training with a new list of slow flashes, and performed a similar check.
The cycle was continued until the number of ``false positives'' reduced
to just a few events. This way, the list of slow flashes for October,
2016, and July, 2017, increased from the initial 798 to 1805 events.
The CNN was trained then with this final list of slow flashes and
applied to the rest of the data resulting in a sample of nearly 8,300
events in comparison with the initial 3,327.

\begin{figure}[!ht]
	\centering
	\mbox{\figh{.25}{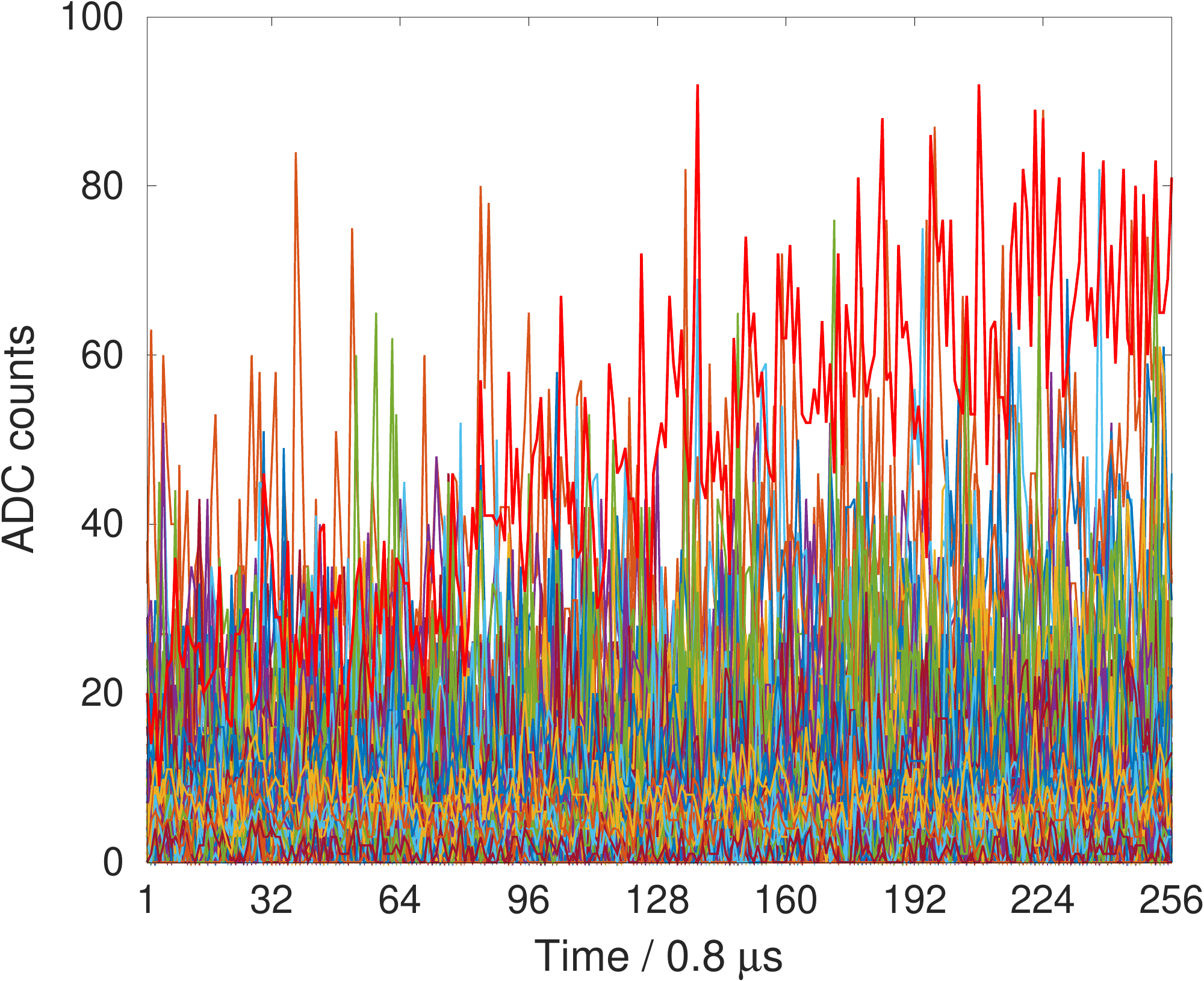}\quad\figh{.25}{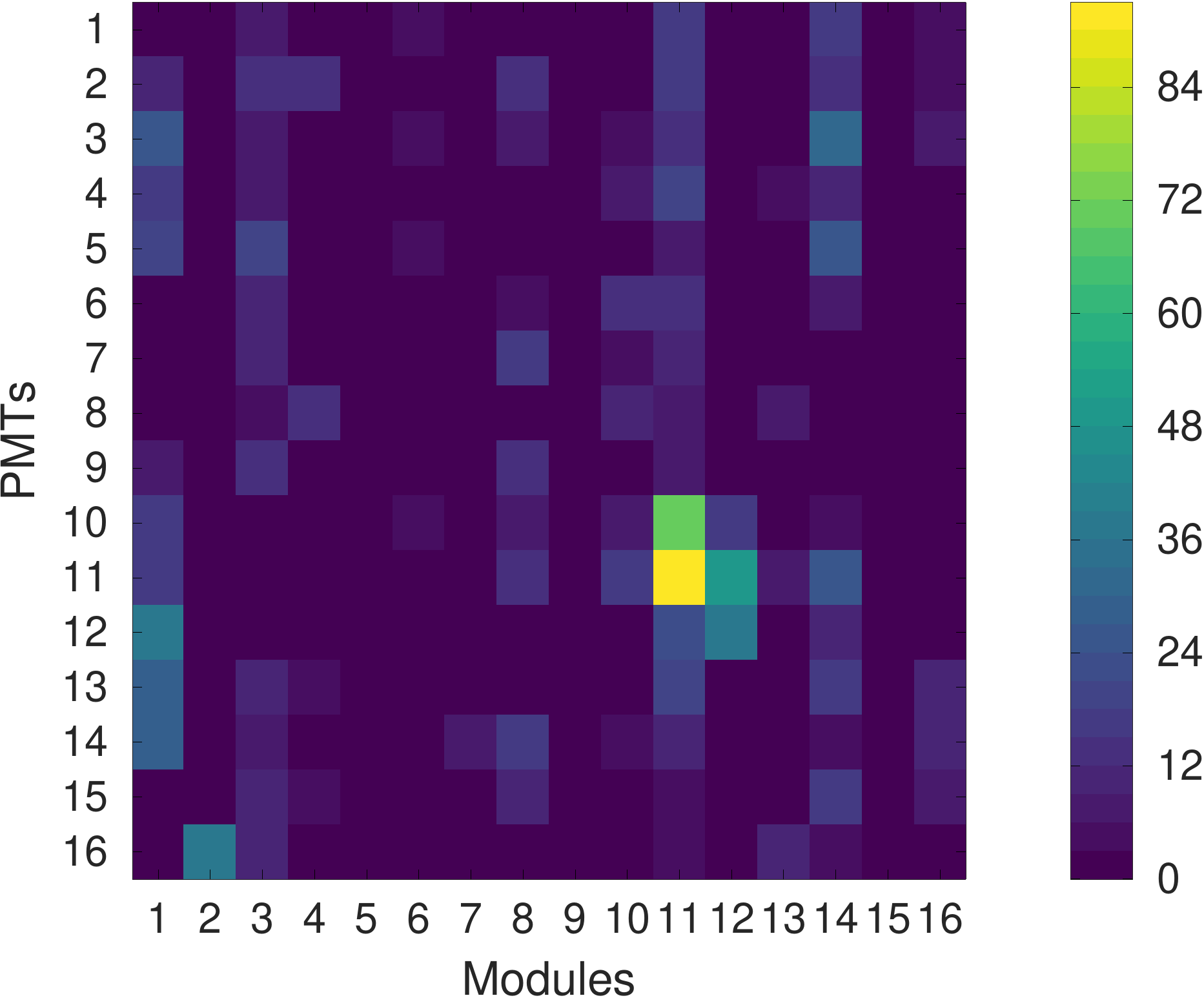}}
	\caption{An event registered above North Canada on November~3, 2016,
	and seemingly mistakenly selected by the CNN as a slow flash.}
	\label{igloolik}
\end{figure}

All newly selected events were checked and more than 100 of them were
excluded from the final sample. More than 30 of them were events with a
weak signal and very few or zero good fits of the waveforms. Eighteen
excluded events represented so called ``steep-front flashes'' mentioned
above that are similar to slow flashes in that they mostly illuminate
the whole focal surface but have waveforms close to TLFs.
The instant growth of the signal in this type of events could not be
identified with input data using only 16 uniformly distributed 
slices of the whole record.
A few events registered on the terminator with low
values of the high-voltage system were excluded, too.
Finally, a few excluded events looked similar to the event
with a non-uniform illumination of the focal surface shown in
Figure~\ref{sf-nonuniform}. One of these events is shown in
Figure~\ref{igloolik}. It was registered on November~3, 2016, above North
Canada. We did not find any information on lightning strikes or
thunderstorms in the region on that date but found that the field of
view of active channels coincides with the location of a small
settlement called Igloolik. The place appears as a small bright spot on
the surrounding dark background in DMSP-OLS Nighttime Lights maps
obtained in the visible and infrared range.\footnote{%
	\url{https://ngdc.noaa.gov/eog/dmsp/downloadV4composites.html}}
Since all events of this kind were registered above populated areas, we
conclude they are likely to have an anthropogenic origin.

\begin{figure}[!ht]
	\centering
	\fig{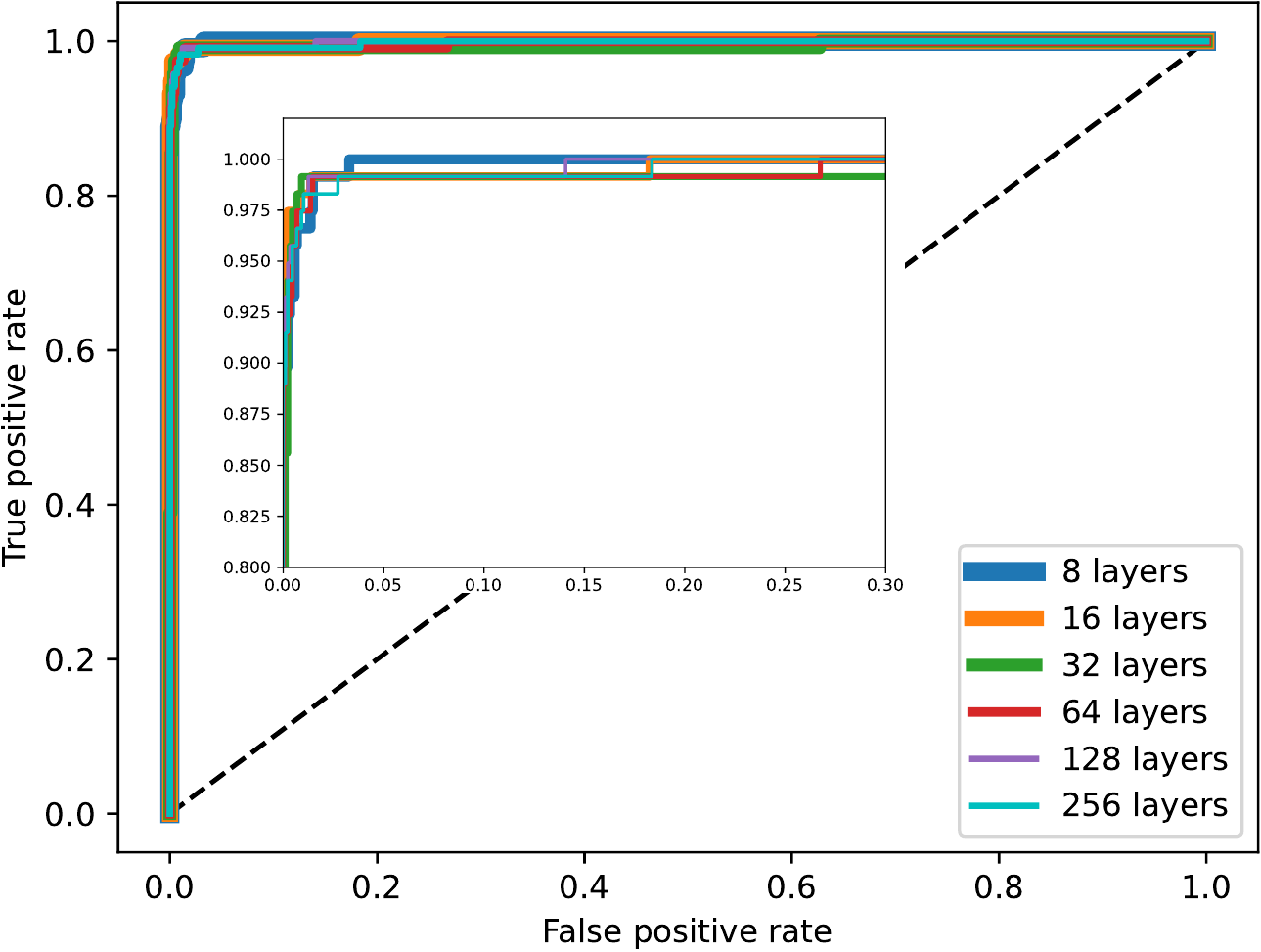}\quad\fig{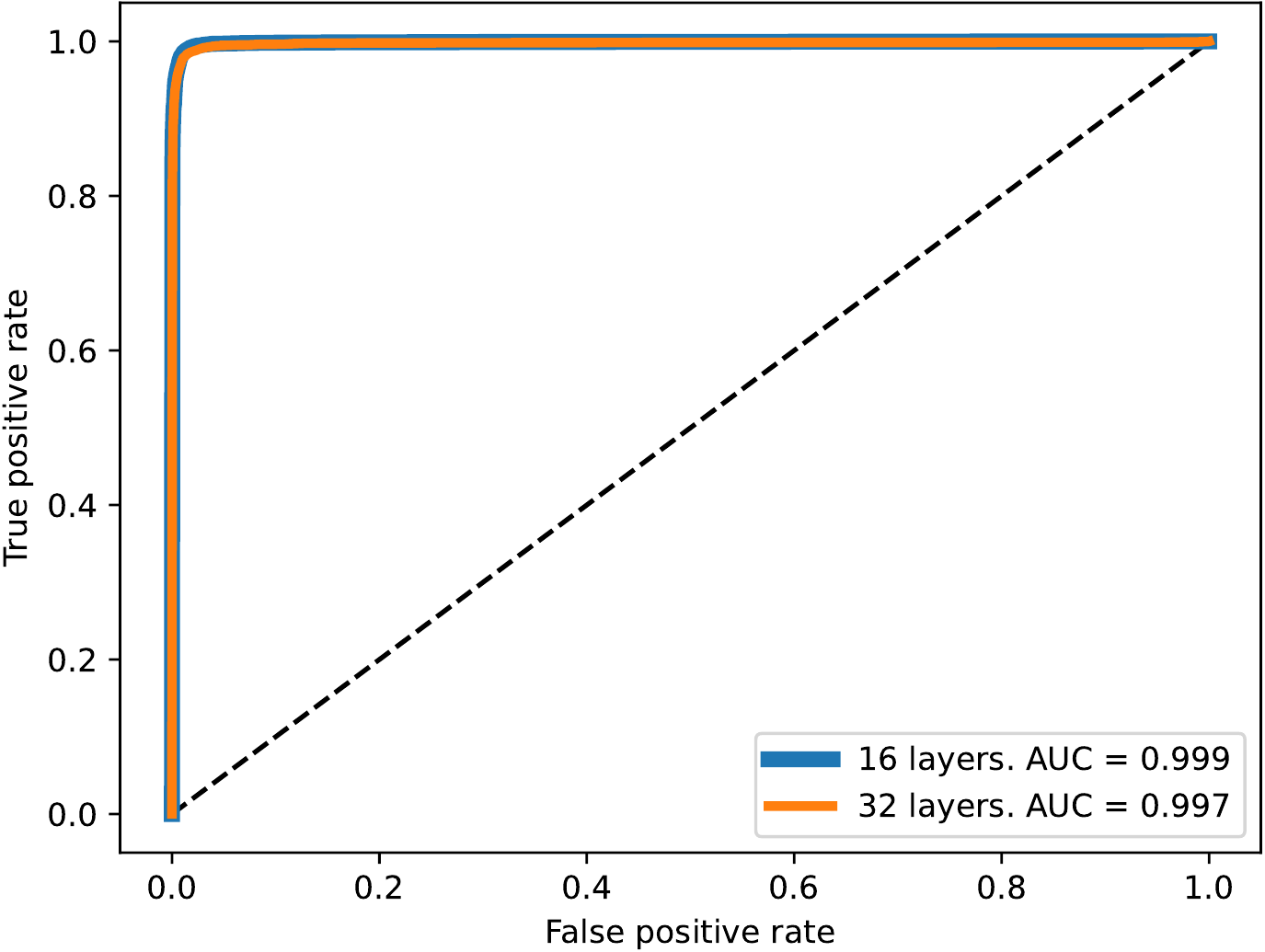}
	\caption{ROC curves for the training (left) and testing (right)
	stages of CNNs with the same architecture but different arrangement
	of input data. See the text for details.}
	\label{sf-roc}
\end{figure}

In the end, we obtained a set of more than 8100 events classified by the
CNN as slow flashes, which was approximately 2.4 times larger than the
initial sample selected with the conventional algorithm.
At this moment we repeated an analysis of performance of CNNs using
different arrangement of the input data in order to verify if the
initial choice of the 16-layer representation was correct.  The left
panel of Figure~\ref{sf-roc}  shows ROC curves for the training stage of
CNNs that employed input data arranged in 8, 16, 32, 64, 128 and 256
layers sampled uniformly from complete records.  It can be seen that
CNNs trained with 8- and 256-layer samples performed slightly worse than
the others but the difference was negligible.  The right panel shows ROC
curves for the testing stage of the CNNs that used 16- and 32-layer
input samples.  The CNN that used simpler data representation (16
layers) performed marginally better than the CNN with 32-layer data (AUC
was equal to 0.999 vs.\ 0.997 respectively).

\begin{figure}[!ht]
	\centering
	\figw{.9}{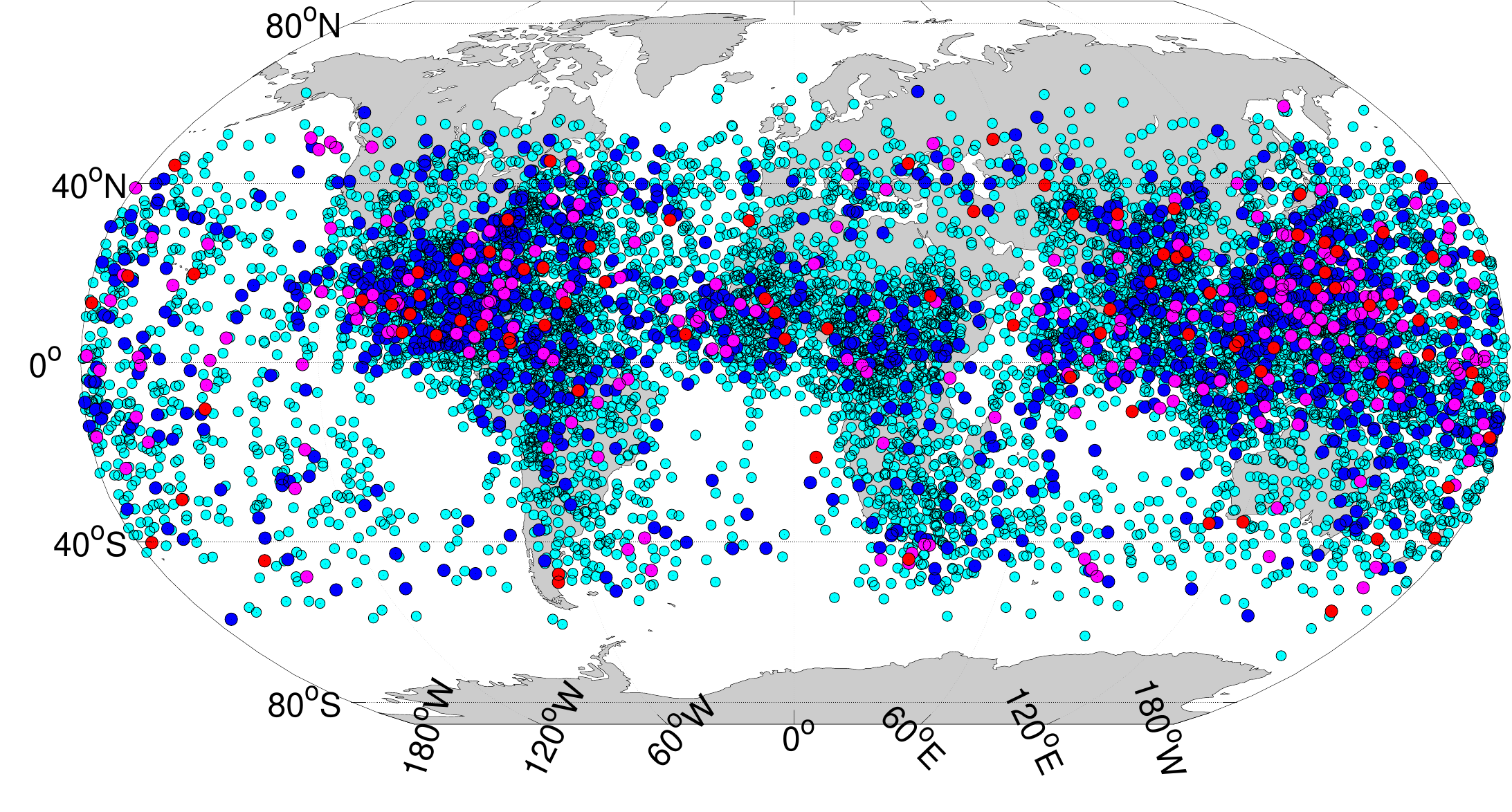}
	\caption{Geographical distribution of slow flashes.
		Colours denote different ranges of amplitudes the same way as in
		Figure~\ref{tlf-map} with the highest ones shown in red and the
		weakest in cyan.}
	\label{sf-map}
\end{figure}

Geographical locations of slow flashes in the final sample are shown in
Figure~\ref{sf-map}. It is clearly seen that they form three huge spots
around the main regions of high thunderstorm activity around the
Carribean, central Africa and South-East Asia. Noticeably, the
distribution of amplitudes of slow flashes is in a sense opposite to
that of TLFs with weak flashes forming the majority of the
sample.

Similar to the case of classifying track-like flashes, the training sets
were imbalanced: slow flashes made up $\sim10\%$ of the samples.
However, a rearrangement of data demonstrated that this did not lead to
a decrease in the accuracy of training.

%______________________________________________________________________________
\section{Discussion}

We presented results of using two types of neural networks---multilayer
perceptrons and convolutional networks---for classification of data
obtained with the TUS orbital fluorescence telescope.  The architecture
of the neural networks was simple and did not put high demands on
computer facilities so that all calculations presented in the work were
performed on a desktop PC without a GPU.  Nevertheless, both MLPs and
CNNs demonstrated a high accuracy in classifying two types of
events registered by TUS: instant track-like flashes produced by cosmic
ray hits of the focal surface of the detector and so called slow flashes
that used to originate from distant lightnings.
More than this, the application of neural networks allowed us to
increase the number of previously known TLFs by approximately 20\% and
the number of slow flashes by 2.4 times due to the discovery of weak
signals omitted by conventional algorithms.

To overcome the incompleteness of marks in the training data sets, we
performed an iterative procedure in which all events selected as false
positives after a run were thoroughly checked and marked as
representatives of the respective class if necessary. The cycle was
repeated until there were no doubts left that the overwhelming majority of
events in the data set were divided into classes properly.  We employed
data from the World-Wide Lightning Location Network for a certain period
of observations to verify the validity of classifying slow flashes by
the neural network.  However, it is necessary to note that the final
samples could contain a number of erroneous records due to the
incompleteness of data on lightning strikes and the fuzzy nature of weak
signals in the TUS data.

The later is worth a special comment. We have already mentioned
that we faced certain problems when trying to classify weak
signals, be it slow flashes or track-like flashes: a visual analysis
of some weak signals did not give a clear clue on how to classify them.
In case of TLFs, one has no way to verify if a weak kink in
ADC codes was caused by a low-energy charged particle hitting a single
PMT or it was just a random fluctuation of the signal.  The only way to
check this to some extent is to perform more detailed simulations of the
photodetector. The situation with slow flashes seems more complicated in
this respect. In principle, one can add information on lightnings to the
classifier.  However, to the best of the author's knowledge, there is
currently no network of lightning detectors uniformly covering the
Earth.  Even in case there was one, it would be difficult to judge
unequivocally if a particular lightning resulted in a slow flash
registered by TUS because of the low accuracy of its timestamps in
comparison with those for lightnings, a wide variety of energies of
lightnings and the diversity of their types. One might expect that
intra-cloud, cloud-to cloud and cloud-to-ground lightnings of the same
power used to produce different illumination of the TUS mirror.  A
detailed simulation seems to be needed to better understand the
phenomenon.  However, the approach looks promising if applied to
lightnings within the FOV of a telescope that has more precise
timestamps.

It was interesting to find out that different architectures of neural
networks and different representations of input data might be needed for
classification of different types of signals within the same data set.
It was also remarkable that there was no need to use complete records of
events for their effective classification. More than this, networks
trained on specially crafted subsets of complete records demonstrated
higher classification accuracy than those employing full data.

We believe that our results help to better understand the data obtained
during the TUS mission. This is important because the TUS data set
provides significant information about the processes taking place in the
nocturnal atmosphere of the Earth in the UV band, which comprise the
background for registering UHECRs by the future full-scale orbital
experiments.  A proper classification of these signals, including weak
ones, will help to reduce the background and possibly to develop
triggers finely tuned for registering UHECRs and suppressing
``parasitic'' signals.  In case of TUS, effective anti-triggers for slow
flashes and instant track-like events could have increased the amount of
``useful'' data by more than 15\%.  This can strongly improve the
efficiency of orbital missions, especially taking into account their
limited hardware and telemetry budgets.  We plan to apply the experience
obtained during the present study for classification of data of the
Mini-EUSO detector currently operating at the ISS and for the
fluorescence telescopes that are under development.

Let us finally remark that there is another type of machine
learning methods that is widely used for classification of data but is
very different from the approach used in the paper. This is the decision
tree and its multiple variants (gradient boosted trees, random forest
etc.).  They are in particular famous because of the simplicity of
understanding and interpreting their results.  We are going to employ
this approach and compare it to the presented results in another paper.

%______________________________________________________________________________
\section*{Acknowledgments}
I thank Margarita~Kaznacheeva for the initial data
set of slow flashes and a helpful discussion.
The research was supported by the Interdisciplinary Scientific and
Educational School of Lomonosov Moscow State University
``Fundamental and Applied Space Research.''
All neural networks were implemented in Python with the Keras
library~\cite{keras} available in TensorFlow~\cite{tensorflow}.

%______________________________________________________________________________
\bibliographystyle{JHEP}
\bibliography{nn4tus}
\end{document}